\documentclass[3p,12pt]{elsarticle}

\usepackage{graphicx, subfigure}
\usepackage[utf8]{inputenc}
\usepackage{amsmath}
\usepackage{amssymb}
\usepackage{units}
\usepackage{booktabs}
\usepackage{eurosym}
\usepackage{threeparttable}
\usepackage{multirow}
\usepackage{natbib}
\usepackage{url}
\usepackage{cprotect}

\journal{Computer Physics Communications}
\begin{document}
\begin{frontmatter}
\title{A Performance Comparison of Different Graphics Processing Units Running Direct $N$-Body Simulations}
\author[sap]{R. Capuzzo--Dolcetta}
\ead{roberto.capuzzodolcetta@uniroma1.it} 
\author[sap]{M. Spera}
\ead{mario.spera@uniroma1.it}
\address[sap]{Dep. of Physics, Sapienza, University of Roma, P.le A. Moro 5, Roma, Italy}

\begin{abstract}
Hybrid computational architectures based on the joint power of Central Processing Units and Graphic Processing Units (GPUs) are becoming popular and powerful hardware tools for a wide range of simulations in biology, chemistry, engineering, physics, etc..
In this paper we present a comparison of performance of various GPUs available on market when applied to the numerical integration of the classic, gravitational, $N$-body problem. To do this, we developed an OpenCL version of the parallel code (\verb+HiGPUs+) used for these tests, because this version is the only apt to work on GPUs of different makes.
The main general result is that we confirm the reliability, speed and cheapness of GPUs when applied to the examined kind of problems (i.e. 
when the forces to evaluate are dependent on the mutual distances, as it happens in gravitational physics and molecular dynamics). More specifically, we find that also the cheap GPUs built to be employed just for gaming applications are very performant in terms of computing speed also in scientific applications and, although with some limitations in central memory and in bandwidth, can be a good choice to implement a machine for scientific use at a very good performance to cost ratio.

\end{abstract}

\end{frontmatter}

\section{Introduction}
The $N$-body problem is the study of the motion of $N$ point-like masses interacting through a pair-wise force that depends only on their positions. It finds its applications 
on both small and large spatial scales. In particular, if we consider astronomical systems, from binary stars up to galaxy clusters, the interaction force is gravity and it is usually possible to neglect relativistic effects. In this case we refer 
to the classical, gravitational, $N$-body problem. \citet{wang91} showed the possibility to give a solution by series of the $N$-body problem, but it results of no practical use due to the exceedingly slow speed of convergence of the solution series expression. Therefore only numerical $N$-body simulations may yield to a deep knowledge of the dynamical evolution of stellar systems.

During the last years both the algorithms and techniques to solve numerically the $N$-body problem and the hardware have been significantly improved. As a result, the number of bodies to integrate has been significantly increased while keeping the total execution time
reasonably small \citep{dehnen11}. 

In broad lines we can group the numerical $N$-body techniques in the following three categories, in dependence on the different ways to evaluate mutual forces: 

\begin{enumerate}
\item \textit{Direct summation} : the force acting on the particle \textit{i} is computed by the complete sum of the contribution due to all the other $N-1$ particles in the system, that is 
\begin{equation}
\mathbf{F}_i=\sum_{\substack{
j=1 \\
j\neq i
}}^{N} G\frac{m_im_j}{r_{ij}^3}\left( \textbf{r}_j-\textbf{r}_i\right) 
\end{equation}
where $m_i$ and $m_j$ are the masses of particles $i$ and $j$, $r_{ij} = \sqrt{\left(x_j-x_i\right)^2+\left(y_j-y_i\right)^2+\left(z_j-z_i\right)^2}$ is the distance between particle $i$ and particle $j$ and $G$ is the gravitational constant. ``Direct summation'' is the simplest to
implement and the most accurate; nevertheless, its computational complexity is high (order of $N^2$), therefore it requires a huge computational power to be successfully applied to big (large $N$) astrophysical systems. The best known codes based on
this approach are \verb+NBODY4+, mainly developed by Sverre Aarseth \citep{aarseth99} ,
$\phi$-\verb+GRAPE+ \citep{harfst07}, 
$\phi$-\verb+GPU+ \citep{berczik11}, 
the \verb+STARLAB+ environment \citep{zwart01}, 
\verb+MYRIAD+ \citep{konsta10},
\verb+NBSymple+ \citep{dolcetta11} and
\verb+HiGPUs+ \citep{dolcetta12}.
\item \textit{Approximation schemes} : the direct sum of inter-particle forces is replaced by another mathematical expression lighter in terms of computational complexity. To this category belongs, for instance, the so called \emph{tree algorithm}, which was originally introduced by \citet{barnes86} and its computational complexity is of $O(N\log N)$. \citet{greengard87}  proposed in the field of molecular dynamics the so called \verb+Fast Multipole Algorithm+, claiming for an $O(N)$ computational time, at least in quasi-homogeneous 2D particles distribution. Unfortunately, the deep comparison between the FMA and the tree code to evaluate gravitational forces performed by \citet{miocchi1998}, showed that FMA has, in 3D, the same $O(N\log N)$ computational complexity of the BH tree code and is slower in both homogeneous and clumpy cases. An example of a modern tree code is \verb+Bonsai+ \citep{bedorf12} but also
\verb+BRIDGE+ \citep{fujii07}
which simultaneously takes advantages from both the direct and the tree approach. Another example is \verb+TreeATD+ (tree-code with Adaptive Tree Decomposition) developed by \citet{miocchi2002}. Another kind of approximation scheme is that developed by \citet{ahmad73}. Using this strategy, during regular steps a direct summation approach is used, but, more frequently, during irregular steps, only the force from neighbour particles is evaluated. The widely used codes \verb+NBODY6+ and \verb+NBODY7+ \citep{nitadori2012} use this scheme.
\item \textit{Grid methods} : many codes are based on the solution of the Poisson's equation
\begin{equation}
\label{poisson}
\nabla^2 \phi (\mathbf{r})=4\pi G \rho(\mathbf{r})
\end{equation}
on a grid leading to a discretized force field (to solve the Poisson's equation, one of the quickest algorithms is the \verb+Fast Fourier Transform+ \citep{fftbook}). This kind of method reduces, as the tree approach, the computational complexity at the expenses of the accuracy. As the tree algorithm, it is widely used in cosmological, large-scale simulations; one of the most known codes which implements the FFT in the solution of the Poisson's equation, in a combination with a tree algorithm, is \verb+GADGET2+ \citep{springel05}.
\end{enumerate}
It is easily understood that the introduction of approximation schemes, in the past years, was compulsory because the computing power needed was too high to use direct summation approaches. All the $N$-body simulations, until $\sim$2006, were made using Central Processing Units (CPUs) or special-purpose machines like the well known \verb+GRAvityPipE+ (\verb+GRAPE+) dedicated boards developed by Sugimoto, Hut and Makino around the end of years 80's \footnote{\url{http://www.astrogrape.org/}}.
Nevertheless, in the last 5-10 years, Graphics Processing Units (GPUs) are slowly replacing CPUs and dedicated hardware for a series of numerical applications because they are getting cheaper and faster while keeping the electric power consumption at very low levels. A detailed discussion about this type of computing evolution can be found in \citep{bedorf12}.
The growth of GPUs as means for scientific computing is strictly linked with the introduction of \emph{Compute Unified Device Architecture} (CUDA\footnote{\url{https://developer.nvidia.com/category/zone/cuda-zone}}, 2006), introduced by the nVIDIA corporation; in fact, thanks to this novelty, nVIDIA graphic cards became easily programmable. CUDA (now at version 5.0) is of simple use because is based on the C programming language, but its limitation is that it can be used to exploit GPUs of the nVIDIA make only. Recently, another GPU programming language has been introduced by the Khronos group: \emph{Open Computing Language} (OpenCL\footnote{\url{http://www.khronos.org/opencl/}}, 2008). OpenCL is based on the programming language C99 and can be used to manage GPUs produced by different vendors (nVIDIA, AMD, etc...) as well as CPUs. Nowadays, although the use of GPUs to accelerate $N$-body codes is widespread, very few codes have been implemented using OpenCL. Therefore, the theoretical computing power of, for example, AMD GPUs has not been fully tested and compared with the performance of nVIDIA GPUs.
This is why, in this paper, we tested the performance of one of our direct summation $N$-body codes called \verb+HiGPUs+ \footnote{Code downloadable at \url{http://astrowww.phys.uniroma1.it/dolcetta/HPCcodes/HiGPUs.html}} which uses OpenCL to exploit and compare the computational power of the modern GPUs available on the market. We will show and discuss the comparison among different GPUs running \verb+HiGPUs+ to evolve various $N$-body systems corresponding to different astrophysical situations, chosen as test cases. Although we have a CUDA version of our code, in this paper we had to use the OpenCL version to compare AMD and nVIDIA GPUs using an identical high-level software. It is worth underlining that the gain of performance running the same test cases on nVIDIA GPUs using the CUDA version of \verb+HiGPUs+, instead of the OpenCL, has been quantified about 5 \%.

Specifically, in Sect. \ref{sec:hardware} we focus our attention to the different hardware (GPUs) tested; in Sect. \ref{sec:perfo} we briefly describe our code and how we measure performance; in Sect. \ref{sec:astromodels} we list the astrophysical test cases, while in Sect. \ref{sec:perform} and \ref{sec:exectimes} we show the results of the performed tests. Finally, in Sect. \ref{sec:conclu} we summarize the results and sketch some future developments.

\section{Hardware}
\label{sec:hardware}
Our performance tests were done on different GPUs manufactured by nVIDIA and AMD corporations. In Tab. \ref{tab:gpu1} and Tab. \ref{tab:gpu2} we list the GPUs used for our benchmarks with some useful reference data.
\begin{table}
\centering

\begin{tabular}{lccccc}
\toprule
\textbf{GPU Model} & \textbf{Launch} & \textbf{Cores} & \textbf{Clock} & \textbf{32bitP} & \textbf{64bitP} \\
 & {\footnotesize \verb+(quarter year)+} & {\footnotesize \verb+(number)+} & {\footnotesize \verb+(GHz)+} & {\footnotesize \verb+(Gflops)+} & {\footnotesize \verb+(Gflops)+} \\
 \midrule
 AMD Radeon HD 6970 & Q4 2010 & 1536 & 0.880 & 2703 & 675 \\
 AMD Radeon HD 7970 & Q1 2012 & 2048 & 0.925 & 3789 & 947 \\
 AMD Radeon HD 7870 & Q1 2012 & 1280 & 1.000 & 2560 & 160 \\
 nVIDIA GeForce GTX 580 & Q4 2010 & 512 &  1.544 & 1581 & 198 \\
 nVIDIA GeForce GTX 680 & Q1 2012 & 1536 &  1.006 & 3090 & 129 \\
 nVIDIA Tesla K20 & Q4 2012 & 2496 & 0.706 & 3520 & 1170 \\
 nVIDIA Tesla C2050 & Q4 2009 & 448 & 1.150 & 1030 & 515 \\
 nVIDIA Tesla C1060 & Q2 2008 & 240 & 1.300 & 622 & 78 \\
 \bottomrule 
\end{tabular}

\cprotect\caption{{\small Some characteristics of each of the tested GPUs. In the column \textit{Launch} the letter \textit{Q} stands for \textit{Quarter}. The columns 32bitP and 64bitP list the maximum theoretical performance in single and double precision respectively in \verb+Gflops+ (billion of floating point operations per second).}}\label{tab:gpu1}
\end{table}
\begin{table}
\centering

\begin{threeparttable}
\begin{tabular}{lccc}
\toprule

\textbf{GPU Model} & \textbf{TDP} & \textbf{Memory} & \textbf{Bandwidth}
\tnote{(a)} \\
 & {\footnotesize \verb+(Watt)+} & {\footnotesize \verb+(MB)+} & {\footnotesize \verb+(GB/s)+} \\
 \midrule
 AMD Radeon HD 6970 & 250 & 2048 & 176.0 \\
 AMD Radeon HD 7970 & 250 & 3072 & 264.0 \\
 AMD Radeon HD 7870 & 175 & 2048 & 153.6 \\
 nVIDIA GeForce GTX 580 & 244 & 1536 & 192.3 \\ 
 nVIDIA GeForce GTX 680 & 195 & 2048 & 192.3 \\
 nVIDIA Tesla K20 & 225 & 5120\tnote{(b)} & 208.0 \\
 nVIDIA Tesla C2050 & 238 & 3072\tnote{(b)} & 144.0 \\
 nVIDIA Tesla C1060 & 188 & 4096 & 102.4 \\
 \bottomrule 
\end{tabular}
\begin{tablenotes}
\item [(a)] {\small \verb+Maximum device to device bandwidth.+}
\item [(b)] {\small \verb+ECC memory supported.+}
\end{tablenotes}
\end{threeparttable}

\caption{{\small Some other relevant data to take into account for each tested GPU. The TDP is the Thermal Design Power which indicates the maximum dissipative power of the cooling system: this is taken as an estimate of the GPU power consumption at full load.}}\label{tab:gpu2}
\end{table} 

For the scopes of this paper, we used 3 GPUs of the AMD Radeon series, 2 GPUs of the nVIDIA GTX series and 3 GPUs of the nVIDIA Tesla series. It is important to stress that while GeForce and Radeon cards are explicitly designed for the gaming market, the Tesla cards are dedicated to scientific users and, since double precision
operations are not needed for playing videogames, both GeForce and Radeon cards have limited 64-bit computing capability. At this regard, we notice from Tab. \ref{tab:gpu1} that the GTX 580 GPU
has a double precision peak limited to 0.125 times its single precision speed, while this ratio is up to 0.5 for the Tesla C2050. Unfortunately, for the last generation card, GTX 680, which is based on the so called Kepler architecture\footnote{\url{http://www.nvidia.com/content/PDF/kepler/NVIDIA-Kepler-GK110-Architecture-Whitepaper.pdf}}, this factor is even smaller (about 0.1). 
Looking at Tab. \ref{tab:gpu2} it is worth noting that both the GPU Tesla C2050, and the most recent Tesla K20, support ECC (Error
Correcting Code) memory which can detect and correct the most common memory errors ensuring, probably, an improved system stability and more reliable results when running very long
simulations. Moreover, Tesla cards have, in general, more on-board memory, up to 5 GB (in some cases 6 GB) in the Tesla K10/K20/K20X. ECC memory, improved 64-bit performance and
large on-board memory are surely important characteristics for scientific users, but these features have an important cost too.
Moreover, basing on declared performance and manufactured characteristics, Radeon GPUs seem to represent a good compromise between 32/64 bit performance, cost and power consumption.
Another feature which emerges from Tab. \ref{tab:gpu1} is that last generation cards have, in general, a greater number of cores with lower operating
frequency than older GPUs. This new ``extreme'' parallel approach, if combined with a better double precision capability, ensures higher theoretical performance both in 32-bit and 64-bit precision, at least in those regimes where the GPU is fully ``loaded''. However, these technical considerations are purely ideal. Actually, the effective measured
performance depends on the combination of hardware, software, drivers, characteristics of the motherboard and many other factors that cannot be taken into account in an easy way. 

In our case, the benchmarks were performed on one of our workstations at the Department of Physics of ``Sapienza'', University of Roma. The main characteristics of this workstation (named
\verb+astroc12+) and the software used to perform our astrophysical test systems are summarized in Tab. \ref{tab:astroc9}.\footnote{The Tesla K20 card has been tested thanks to two remote accesses kindly provided by Simon Portegies Zwart, to a machine sited at the Department of Astronomy of the Leiden University (NL), and by the E4 computer engineering to one of their test workstations.}
\begin{table}
\centering

\begin{threeparttable}
\begin{tabular}{ll}
\toprule
\multicolumn{2}{c}{\textbf{Astroc12 workstation characteristics}}\\
\midrule
Motherboard & ASUS P6T7 WS SuperComputer\\
Power Supply & Enermax ERV1250EGT 1250 W \\
CPU & 1 Intel core i7 950 @ 3.07 GHz \\
RAM memory & 6 GB (3 x 2GB) 1333 MHz \\
Operating System & Ubuntu Lucid 10.04.2 64-bit version \tnote{(a)} \\
OpenCL & AMD and nVIDIA implementations version 1.2 \tnote{(b,c)}\\
CUDA & version 4.0, May 2011 \tnote{(c)} \\
AMD Drivers & Catalyst 12.6 Linux x86\_64 \tnote{(b)} \\
nVIDIA Drivers & 295.75 Linux x86\_64 \tnote{(c)} \\
Compiler & gcc/g++ version 4.4.4 \\
MPI & OpenMPI version 1.5.4 \tnote{(d)} \\
GPU & see Tab. \ref{tab:gpu1} \\
Software used & \verb+HiGPUs+ (direct $N$-Body code) \\
\bottomrule 
\end{tabular}
\begin{tablenotes}
\item [(a)] {\small \url{http://www.ubuntu.com/}}
\item [(b)] {\small \url{http://developer.amd.com/zones/OpenCLZone/Pages/default.aspx}}
\item [(c)] {\small \url{http://developer.nvidia.com/category/zone/cuda-zone}}
\item [(d)] {\small \url{http://www.open-mpi.org/}}
\end{tablenotes}
\end{threeparttable}

\caption{{\small The main characteristics of our workstation used to benchmark the GPUs listed in Tab. \ref{tab:gpu1} and  Tab. \ref{tab:gpu2}}}\label{tab:astroc9}
\end{table}

\section{Performance measurements}
\label{sec:perfo}
The benchmark tests were done with our new, direct summation $N$-body code called \verb+HiGPUs+ \citep{dolcetta12}. It implements an Hermite, 6th order, time integration algorithm, consisting essentially in a predictor-evaluation-corrector scheme, with block time steps as proposed in \citep{nitadori08}. The ``block time stepping'' is a technique which allows each particle to have its own time step, which is determined using a properly modified version of the generalized Aarseth criterion \citep{nitadori08} and approximated to the nearest power of two. In this way, particles are organized in groups (blocks) which share the same time step. Objects with smaller time steps must be updated more often than those with larger time steps, whose positions and velocities are predicted but not updated (corrected) through the evaluation of their accelerations and higher order time derivatives. In particular, if we call, hereafter, $m$ the number of particles to update in one integration step, the computational complexity of the problem is reduced from $O(N^2)$ to $O(mN)$ where $m$ gets equal to $N$ at the end of the time synchronization process. 
The \verb+HiGPUs+ code combines tools of C and C++ and uses CUDA or OpenCL coupled with MPI and OpenMP to fully exploit the power of hybrid computing platforms (CPUs+GPUs). A detailed description of the code implementation and its scalability is found in \citep{dolcetta12}.
Here we limit to list in Tab. \ref{tab:sections} a schematic representation of our code, divided, for convenience, in 11 operational sections. In this paper we study the GPU performance for sections 2, 3, 4, 5, 6, 7, 10 and 11, measuring the time to complete each of them deducing the speed in \verb+Gflops+.

In broad lines, our strategy to measure performance can be summarized by the following statement: if, for the $k$-th section, the total number of running GPU threads is $T_k$, the counted floating point operations are $F_k$ and the time to complete the section, in seconds, is $\Delta t_k$, the performance $R_k$ in
\verb+Gflops+ is obtained by the formula 
\begin{equation}
\label{getperf}
R_k = \frac{T_k F_k}{10^9 \Delta t_k}.
\end{equation}

\begin{table}

\begin{threeparttable}
\begin{tabular}{cccc}
\toprule
\textbf{Section} & \textbf{Description} & \textbf{Data for measuring} \\
& & \textbf{performance}\\
\midrule 
 & Each node determines the stars to be updated and &  \\
1 & their indexes are stored in an array named \emph{next} & Not used\tnote{(a)} \\
 & containing \emph{m} integer elements &  & \\
\midrule
2 & Each node copies to its GPUs the array containing & $4m$ \verb+Bytes+\\
& indexes of \emph{m} particles \\
\midrule
 & If $k$ is the number of GPUs that will be used & \\
3 & in the numerical integration, the predictor step of & \verb+81 ops (DP)+\\
 & of \emph{N/k} stars is executed &  \\
\midrule
 & Each node computes & \verb+82 ops (SP)+\\
4 & the forces (and their higher order derivatives) of \emph{m} & \verb+15 ops (DP)+ \\
 & particles due to \emph{N/k} bodies  & \\
\midrule
5 & Each node reduces the calculated forces and derivatives & \verb+10 ops (DP)+\\
 & of \emph{Bfactor} blocks & \\
\midrule
6 & Each node adjusts conveniently the reduced values & \verb+ 32 BR+\\
& & \verb+32 BW+\\
\midrule
7 & The CPUs receive the accelerations from the GPUs & $96m$ \verb+Bytes+\\
\midrule
8 & The \verb+MPI_Allreduce()+ functions collect and & Not used\tnote{(a)} \\
 & reduce accelerations from all the computational nodes & \\
\midrule
9\tnote{(b)} & Corrector step and time step update for \emph{m} stars & Not used\tnote{(a)}\\
\midrule
 & The reduced accelerations (and derivatives) and & \\
10\tnote{(c)} & the corrected positions and velocities of \emph{m} bodies are &  $192m$ \verb+Bytes+\\
 & passed to the GPUs of each node & \\
\midrule 
11\tnote{(c)} & The GPUs rearrange the updated particles following & \verb+36 BR+ \\
 & the original indexes stored in the array \textit{next} & \verb+24 BW+\\
 \bottomrule
\end{tabular}
\begin{tablenotes}
\item [(a)] {\small \verb+This section involves only the CPU.+}
\item [(b)] {\small \verb+This section has been ported on GPU in the latest version of HiGPUs.+}
\item [(c)] {\small \verb+This section is not needed if the corrector step is performed on the GPU.+}
\end{tablenotes}
\end{threeparttable}

\caption{{\small The main sections of our code performed for each time step. The \textquotedblleft{}convenient
adjustment\textquotedblright{} mentioned in the 6th section of our
code is referred to the reorganization of the computed and reduced
accelerations and derivatives in only one array (instead of three)
to improve the performance of the following data transfer from the
GPU to the CPU. In this way we execute one bigger copy instead of
three smaller. The notation DP stands for double precision while SP 
for single precision. The notation BR stands for Bytes read and BW 
for Bytes written per GPU thread, from and to the GPU global memory. With the word \textit{Bfactor} 
we indicate a variable introduced in our code to split further the computation among the GPU 
threads which is useful especially when the number of particles
to update is small. For all the details, which goes beyond the target of this paper, see \citep{dolcetta12}}.}
\label{tab:sections}
\end{table}
 
To count the floating point operations we refer to Tab. \ref{tab:flops} which has been built following the Table 5-1 of the CUDA C programming guide\footnote{\url{http://docs.nvidia.com/cuda/pdf/CUDA_C_Programming_Guide.pdf}}
and the information given in the whitepaper of the FERMI architecture\footnote{\url{http://www.nvidia.com/content/PDF/fermi_white_papers/NVIDIA_Fermi_Compute_Architecture_Whitepaper.pdf}}.
Moreover, it is important to underline that \verb+HiGPUs+ uses both single and double precision arithmetic in the main GPU kernel (evaluation of mutual forces). Double precision is used to calculate inter-particle distances and to cumulate accelerations and their higher order time derivatives, while all the other operations are executed in single precision. This approach is adopted to take advantage of the higher speed of GPUs in performing 32-bit operations, but, at the same time, it keeps a sufficiently high precision in the force evaluation and, so, in the time integration. Anyway, also other approaches are found in the literature. For example, the use of emulated double precision or pseudo-double precision, (also called Double-Single, DS, precision) is widely used (see, for example \citep{aarseth99} and \citep{berczik11}).
In this way, only single precision operations are performed, replacing a 64-bit value with two, properly handled, 32-bit values. We have implemented a DS version of \verb+HiGPUs+ but we notice that, although the performance is higher in terms of \verb+Gflops+ (especially for nVIDIA GeForce GPUs), the particle time steps distribution exhibits a sort of tail in the area of small time steps which is not present when using double precision to evaluate accelerations and higher order derivatives (this peculiar behaviour has already been pointed out by \citet{gaburov09} comparing single and double-emulated precision).
 Therefore, using the DS version of \verb+HiGPUs+, we obtain higher performance but a total execution time which is the same or greater and a relative energy conservation which is, on average, 2 orders of magnitude worse. In fact the generalized Aarseth criterion \citep{nitadori08} used by \verb+HiGPUs+ to determine the time steps is sensible to round off errors in the calculation of higher order derivatives thus producing the above-said tail.
We do not discuss further this point here because it is out of the scopes of this paper, although a better investigation of this behaviour will likely lead to fix it in a future implementation of our code. 

In Tab. \ref{tab:sections}, sections 2, 7 and 10 involve memory transfers between the GPU and the CPU through the PCI Express interface. Table \ref{tab:sections} shows also the total amount of data, in bytes, that must be exchanged.
On the other hand, sections 6 and 11 involve only read and write operations inside the GPU on-board memory. This is why, for these sections, we measured the execution times in seconds and we give an estimate of the device-to-device memory bandwidth exploited. Table \ref{tab:sections} lists, also, the number of bytes that each GPU thread must read (BR)/write (BW) from/to the GPU memory.

\begin{table}
\centering

\begin{tabular}{ccc}
\toprule
\textbf{Operation} & \textbf{CUDA/OpenCL expression} & \textbf{Equivalent fl. ops.} \\
\midrule
$a\pm b$ & \verb+a+ $\pm$ \verb+b+ & 1 \\
$a\cdot b$ & \verb+a * b+ & 1 \\
$1/\sqrt{a}$ & \verb+rsqrt(a)+ & 4 \\ 
$a/b$ & \verb+a / b+ & 5 \\
$a^b$ & \verb+pow(a,b)+ & 9 \\
\bottomrule 
\end{tabular}

\caption{{\small Number of floating point operations corresponding to the main arithmetical and fundamental operations. The quantities $a$ and $b$ are two generic floating point numbers.}}\label{tab:flops}
\end{table}

\section{Astrophysical models}
\label{sec:astromodels}
The astrophysical models chosen for our tests include low$-N$ cases (256 stars) up to high$-N$ systems (262,144 stars) and their main parameters are listed in Tab. \ref{tab:models}. The first three models refer to systems containing bodies randomly distributed in a sphere of unitary radius. The values of $N$ are 256, 512 and 1,024 starting from an initial ``cold'' condition, i.e. the case where the virial ratio  
\begin{equation}
\label{virialratio}
Q=\frac{2T}{\left| \Omega \right|}
\end{equation}
is equal to zero, where $T$ is the kinetic energy and $\left | \Omega \right|$ the absolute value of the potential energy $\Omega$, given by
\begin{equation} \label{eq:kin_pot}
T=\sum_{i=1}^{N}\frac{1}{2}m_iv_i^2 \qquad
\Omega=-\frac{1}{2}\sum_{\substack{
\left(i,j\right) =1 \\
i\neq j
}}^{N} G\frac{m_im_j}{r_{ij}}.
\end{equation}
In Eq. \ref{eq:kin_pot},  $v_i$ is the speed of the $i$-th star and $r_{ij}$ is the distance between particles $i$ and $j$.
For the masses of the stars, we assumed a bimodal distribution containing $N/2$ ``light'' particles of mass $m_l$ and $N/2$ ``heavy'' particles of mass $m_h$.
We also considered the presence of an external force-field by mean of a time-independent Plummer potential \citep{plummer1911}
\begin{equation}
\label{eq:plumpot}
\phi(r)= \frac{GM_g}{\sqrt{r^2+b^2}},
\end{equation}
where $r$ is the distance to the system barycentre, $b$ is a scale radius and $M_g$ is the total gas mass. In the hypothesis that the external potential mimics the role of a gas residual after star formation, the value of $M_g$ is determined by assuming a value for the Star Formation Efficiency, defined by 
\begin{equation}
\epsilon=\frac{M_*}{M_*+M_g},
\end{equation}
where $M_*$ is the total mass in stars. Here we take $\epsilon=0.3$ as a likely astrophysical value.

On this basis we define three simple reference models, indicated with the symbols \verb+V1+, \verb+V2+ and \verb+V3+, to (roughly) mimic the initial state of young and very young open clusters which are observed in sub-virial conditions, mass segregated, despite their age, and still embedded in their native gas. A deep scientific analysis of the results obtained from these simulations will be  presented elsewhere \citep{dolcetta13} while here we limit the analysis to the GPUs performance.

We also sampled the initial conditions for other $N$-body systems from two King models \citep{king66}, indicated in Tab. \ref{tab:models} with \verb+K1+ and \verb+K2+, with $N_{K1}=65,536$, $N_{K2}=32,768$. For the King models we assumed two values for the dimensionless central concentration parameter, $W_{0_{K1}}=7$ and $W_{0_{K2}}=9$. In model \verb+K1+ an Initial Mass Function, like that described in \citep{kroupa01}, has been adopted while in the model \verb+K2+ all the stars have the same mass. We also sampled a King model, indicated with the letter \verb+K3+, with $N_{K3}=262,144$ stars, $W_{0_{K1}}=6$ and the same mass function used for the model \verb+K1+, embedded in a rough representation of the Milky Way bulge potential as a Plummer analytical potential \citep{allen91} and moving on a circular orbit at 2 kpc from the centre of the system barycentre. We also sampled a Plummer model, indicated with \verb+P1+, with $N_{P1}=16,384$. In the Plummer model all the stars have the same mass.

All these models were generated using \verb+McLuster+ \citep{kupper11} and all the mentioned test cases have been followed up to 10 time units, which is an extension in time sufficient to obtain a reliable averaged performance for all the different sections of our $N$-body code.

\begin{table}
\centering

\begin{threeparttable}
\begin{tabular}{lccccc}
\toprule
\textbf{Model} & \textbf{Notation} & \textbf{N} & \textbf{System} & \textbf{Background}  \\
    &   &    &  \textbf{parameters}  &  \textbf{parameters} \\
\midrule
Homogeneous sphere + & \verb+V1+ & 256    & $R=1$        & $b=1$       \\
Plummer background   &           &        & $M=0.3$      & $M_g=0.7$   \\
\midrule
Homogeneous sphere + & \verb+V2+ & 512    & $R=1$        & $b=1$       \\
Plummer background   &           &        & $M=0.3$      & $M_g=0.7$   \\
\midrule
Homogeneous sphere + & \verb+V3+ & 1,024  & $R=1$        & $b=1.0$     \\
Plummer background   &           &        & $M=0.3$      & $M_g=0.7$   \\
\midrule
Plummer sphere       & \verb+P1+ & 16,384 & $b=1$        &  no         \\
                     &           &        & $M=5$        &  background \\
\midrule
                     &           &        & $W_0=9$      &             \\
King distribution    & \verb+K2+ & 32,768 & $r_c=0.2$    &  no         \\
                     &           &        & $M=5$      &  background \\
\midrule
                     &           &        & $W_0=7$      &             \\
King distribution    & \verb+K1+ & 65,536 & $r_c=0.2$    &  no         \\
				     &           &        & $M=5$      &  background \\
\midrule
King distrib. in a   &           &        & $W_0=6$      &  $b=4$      \\
Plummer background   & \verb+K3+ & 262,144& $r_c=0.01$   &  $M_g=14$   \\
                     &           &        & $M=0.001$    &             \\                      
\midrule
 \bottomrule 
\end{tabular}
\end{threeparttable}

\caption{{\small The complete set of simulations performed for our benchmarks. $R$ and $M$ represent, respectively, radius and mass of the stellar system. The parameter $b$ is the Plummer's core radius (see Eq. \ref{eq:plumpot}), $M_g$ is the total mass of the analytic, stationary background, if present, $r_c$ is the King's core radius and $W_0$ is the dimensionless central concentration \citep{king66}. All the simulations are performed in units such that $G=1$, while the length and mass units are chosen for computational convenience as in column 4.}}\label{tab:models}
\end{table}

\section{Performance results}
\label{sec:perform}

\subsection{Evaluation of the mutual forces}
First of all we analyse the most important section of any $N$-body code: the evaluation of the accelerations and, for the Hermite 6th order scheme, some of their time derivatives. For populous stellar systems this is, by far, the section which takes most of the execution time, therefore the performance exhibited in this part is of crucial importance for realistic scientific applications. On the other hand, for small$-$ and intermediate$-N$ systems ($N \lesssim 16k$), as we will see later, the time spent to execute this evaluation step becomes comparable to (or even smaller than) that spent to complete other \verb+HiGPUs+ sections. This underlines the importance to have powerful and efficient hardware also on small scales. The computational complexity of the evaluation section is  of $O(mN)$. This means that the overall performance of a generic GPU, evolving a system containing $N$ bodies for a certain interval of time, depends strictly on its speed for $m=<m>$ where 
\begin{equation}
<m>=\frac{\sum_{i=1}^{S}m_i}{S}
\end{equation}
where $S$ is the total number of integration steps and $m_i$ the number of particles to update in the $i-th$ step (\citet{berczik11}). The value of $<m>$ depends on many factors like the choice of initial conditions, the value of the softening parameter\footnote{The softening parameter $\epsilon$ is a constant which smooths close gravitational encounters by substituting the inter-particle distance $r_{ij}$ with the expression $\tilde{r}_{ij}^2=r_{ij}^2+\epsilon^2$}, the criterion used for determining the particle time steps, the presence of a very massive body (black hole), the used time integration algorithm other than on the precision used in the integration (double or single). Determining and discussing the relation between $<m>$ and $N$ is out of the scope of this work but we must underlining that the Figures that we are going to show in this Section cannot be taken as indicators of sustained performance but only of the speed to evaluate forces for a particular value of $m$ (problem of computational complexity of $O(mN)$). To know the real performance and gain of a specific GPU on the overall evolution (10 time units for this work) the Figures shown in this Section must be viewed together with the histograms in Fig. \ref{fig:isto} that represent the measured wall-clock times to evolve each test system on each tested GPU.
\subsubsection{large N case: systems K3 and K1}
\begin{figure}[!ht]
\centering
\includegraphics[width=0.8\textwidth]{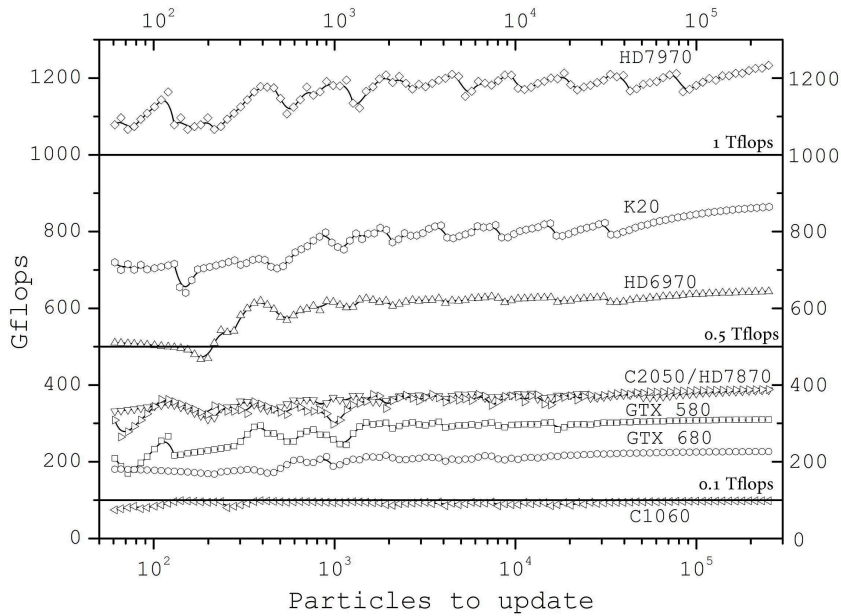}
\cprotect\caption{Speed performance of the tested GPUs, in \verb+Gflops+, as a function of the number of particles to update. This figure refers to system K3, with $N$=262,144.}
\label{fig:highn}
\end{figure}

In Fig. \ref{fig:highn} we show the speed performance of the various GPUs in the execution of the evaluation step of \verb+HiGPUs+, in function of the number $m$ of particles to be updated, in a generic time step. We refer only to the system \verb+K3+ because the resulting plot for system \verb+K1+ does not point out significant differences.
Fig. \ref{fig:highn} shows that, in the whole range of values of $m$, the Radeon card HD7970 performs over 1 \verb+Tflops+ while the other GPUs show a speed from 10\% (Tesla C1060) to $\sim$50\% (Radeon HD6970) up to 75\% (Tesla K20) that of the HD7970. Tesla C2050 and HD7870 have approximatively the same performance while GTX cards do not get considerable results mainly because of the low double precision computing power (see Tab. \ref{tab:gpu1}). Moreover, the new generation card of the GTX class, the GTX 680 (Kepler Architecture), has a speed performance a factor 1.4 worse than that of the previous generation GTX 580 (Fermi Architecture). This is mainly due to the ratio of performance in 64 bit precision operations between these two cards. Nevertheless, it is curious to highlight that, although the technical features of the GTX 680 and HD7870 are very similar, the performance of the HD7870 is, in this large-$N$ regime, about a factor 1.6 higher of GTX 680. This is likely due to that an HD7870 can run up to 51,200 GPU threads in parallel while a GTX 680 only up to 16,384. Therefore, the high parallel capability of the HD7970 is clearly preferable in regimes of full load state of the GPU (as happens in the large-$N$ case). 
Despite tuned and different optimizations introduced in our code working when $m$ is smaller than the maximum number of parallel threads that a GPU can run simultaneously, we can see a slight decay of performance when this number is less than 400 for system \verb+K3+ (700 for system \verb+K1+) especially in the case of the Radeon HD7970, HD6970 and Tesla K20. This is not surprising because these three cards are massively parallel. These GPUs have a large number of processing elements with low clock frequencies and an HD7970 can run up to 81,920 threads simultaneously while an HD6970, as well as a Tesla K20, up to 32,768. Therefore, it is difficult to load fully these GPUs in the above low $m$ regime while the others GPUs are easier to exploit having, in general, both less resources available and less theoretical computing power. This explains why the performance decay at low $m$ of these GPUs is almost negligible. In any case, we can affirm that, for a direct $N$-body code and, more generally, for a kernel which fully loads the GPU using both single and double precision operations, an HD7970, an HD6970 or a Tesla K20 is the best choice to obtain scientific results in a short time.

\subsubsection{intermediate-N case: systems P1 and K2}
\begin{figure}[!ht]
\centering
\includegraphics[width=0.8\textwidth]{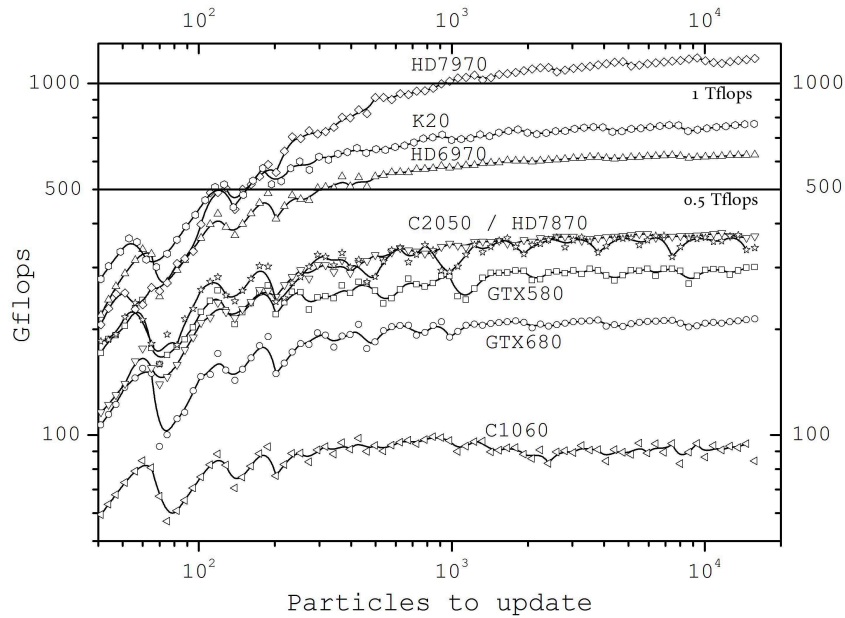}
\cprotect\caption{Speed performance of the tested GPUs, in \verb+Gflops+, as a function of the number of particles to update. This figure refers to system P1, with $N$=16,384.}
\label{fig:mediumn}
\end{figure}

\begin{figure}[!ht]
\centering
\includegraphics[width=0.8\textwidth]{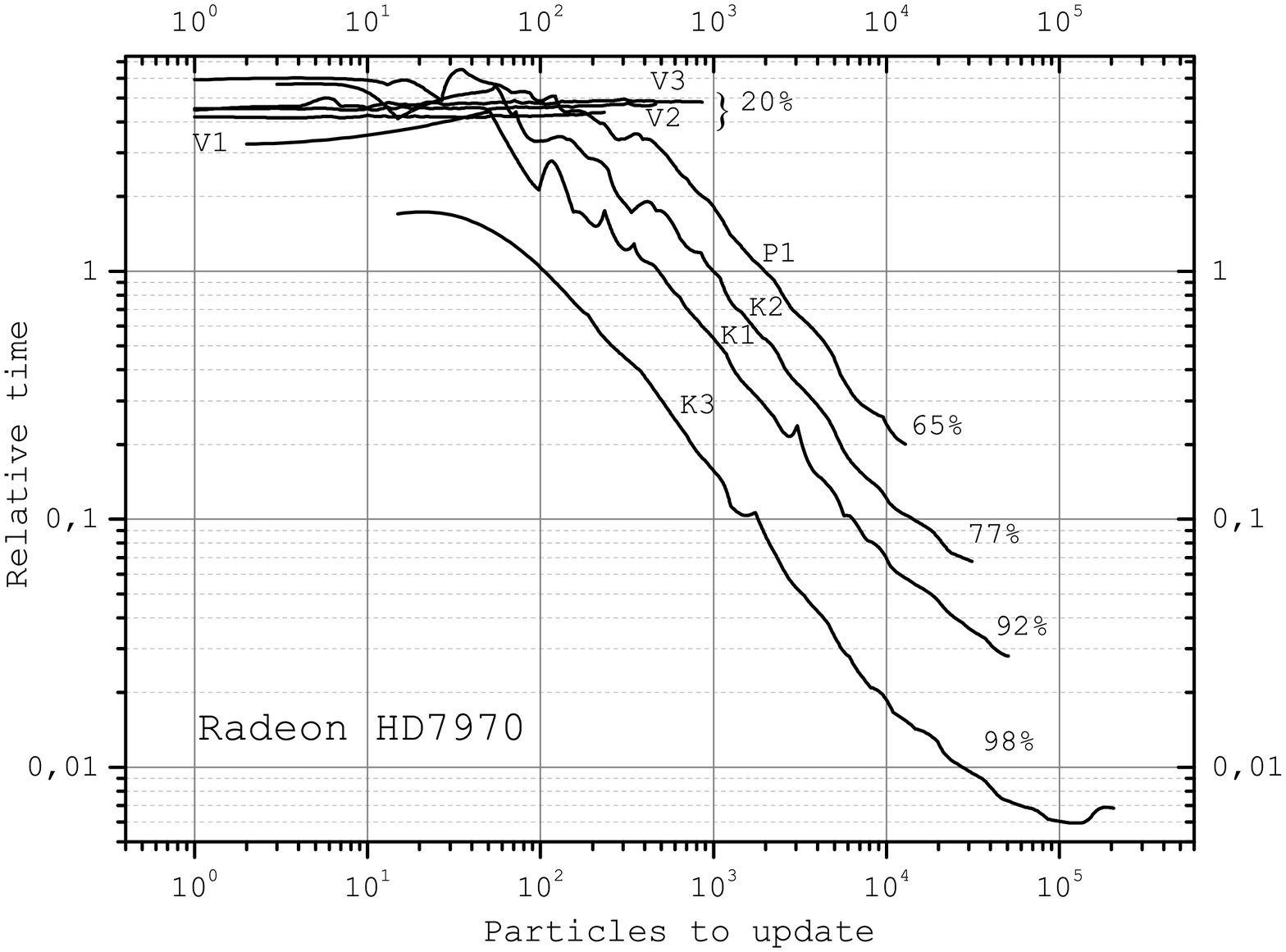}
\caption{Relative importance of all the code sections (excluding the evaluation) to the evaluation section in function of the number of particles to update in different cases. The various curves are labelled by the percentage time spent in the evaluation. }
\label{fig:fig3}
\end{figure}

In Fig. \ref{fig:mediumn} we show the performance of the tested GPUs in the same frame adopted for Fig. \ref{fig:highn}. This figure refers to the system \verb+P1+,  chosen as reference case for this regime of intermediate $N$. Radeon cards, also in this regime, exhibit higher performance than the other GPUs with reference to the evaluation step of \verb+HiGPUs+. The performance of the Tesla K20 remains always between the two Radeon GPUs, except for low values of $m$ ($m\lesssim 200$) in which it performs slightly better. The gain of the massively parallel cards is relevant when $m\gtrsim 500$ while for smaller values of $m$ the performance decay is more evident than in the previous high-$N$ case for all the GPUs although the Tesla C1060 remains in a state of full load (around 80 \verb+Gflops+) because it has, both, less cores and much a lower theoretical performance than the other cards. We found that system \verb+P1+ ($N$=16,384) is a lower limit for the number of particles per GPU in the sense that below this $N$ the time spent by, for example, an HD7970, to complete the other sections of our code becomes significant if compared to the total execution time. 
To stress this idea, Fig.  \ref{fig:fig3} shows the ratio between the sum of the times spent by an HD7970 to complete all the other parts of \verb+HiGPUs+ and that to complete just the evaluation step, as a function of the number of particles to update for our test systems. The fraction of the time spent to evaluate accelerations to the total execution time is about 65\% for system \verb+P1+ using an HD7970. The remaining 35\% is equally divided in memory transfers and reduction of partial forces. Therefore, while for systems \verb+K1+ and \verb+K3+ the evaluation step is by far the most important part, this is no longer true for the other test systems. Fig. \ref{fig:fig3} is useful to show that for systems with $N \lesssim 16k$ the overall hardware performance is determined also by the other sections of \verb+HiGPUs+.

\subsubsection{Low-N cases: systems V1, V2 and V3}
Even if, in this regime, one may not need to use powerful computing accelerators because the total execution time is limited well below that spent to integrate systems in the intermediate and large-$N$ cases, it is very interesting to study how GPUs perform when they are not loaded in full. This may also give us some general and useful information in the case when more than one computing node is available. In fact, for example, a system of $N=1,024$ bodies on a single GPU can be considered almost equivalent to a system composed by $N=1M$ bodies distributed over 1,024 GPUs. Therefore, considering the low-$N$ regime, we can argue some considerations about the performance that would be got running large-$N$ systems over a set of GPUs.

\begin{figure}[!ht]
\centering
\includegraphics[width=0.8\textwidth]{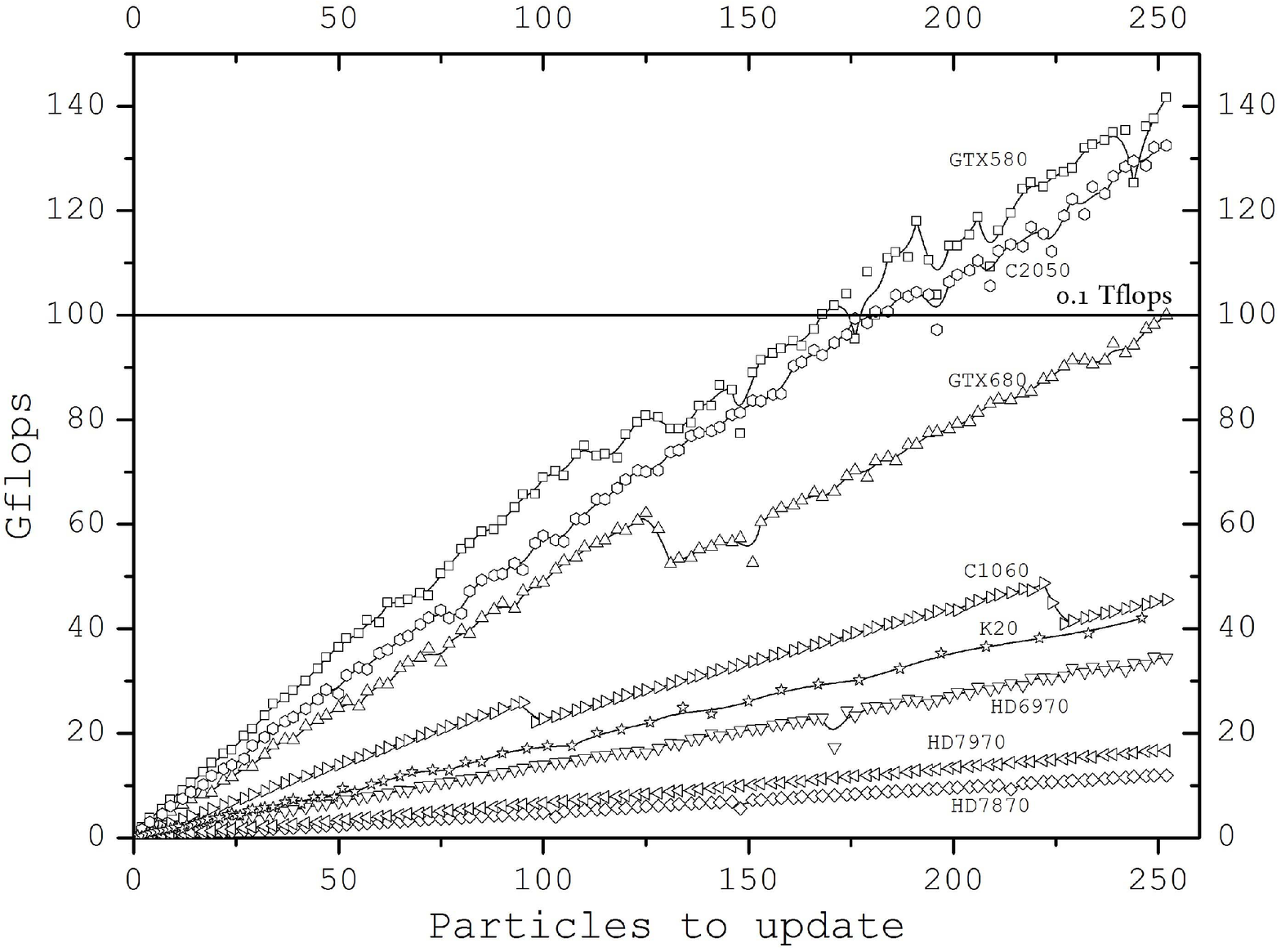}
\cprotect\caption{Speed performance of the tested GPUs, in \verb+Gflops+, as a function of the number of particles to update. This figure refers to system V1, with $N$=256.}
\label{fig:fig4}
\end{figure}

\begin{figure}[!ht]
\centering
\includegraphics[width=0.8\textwidth]{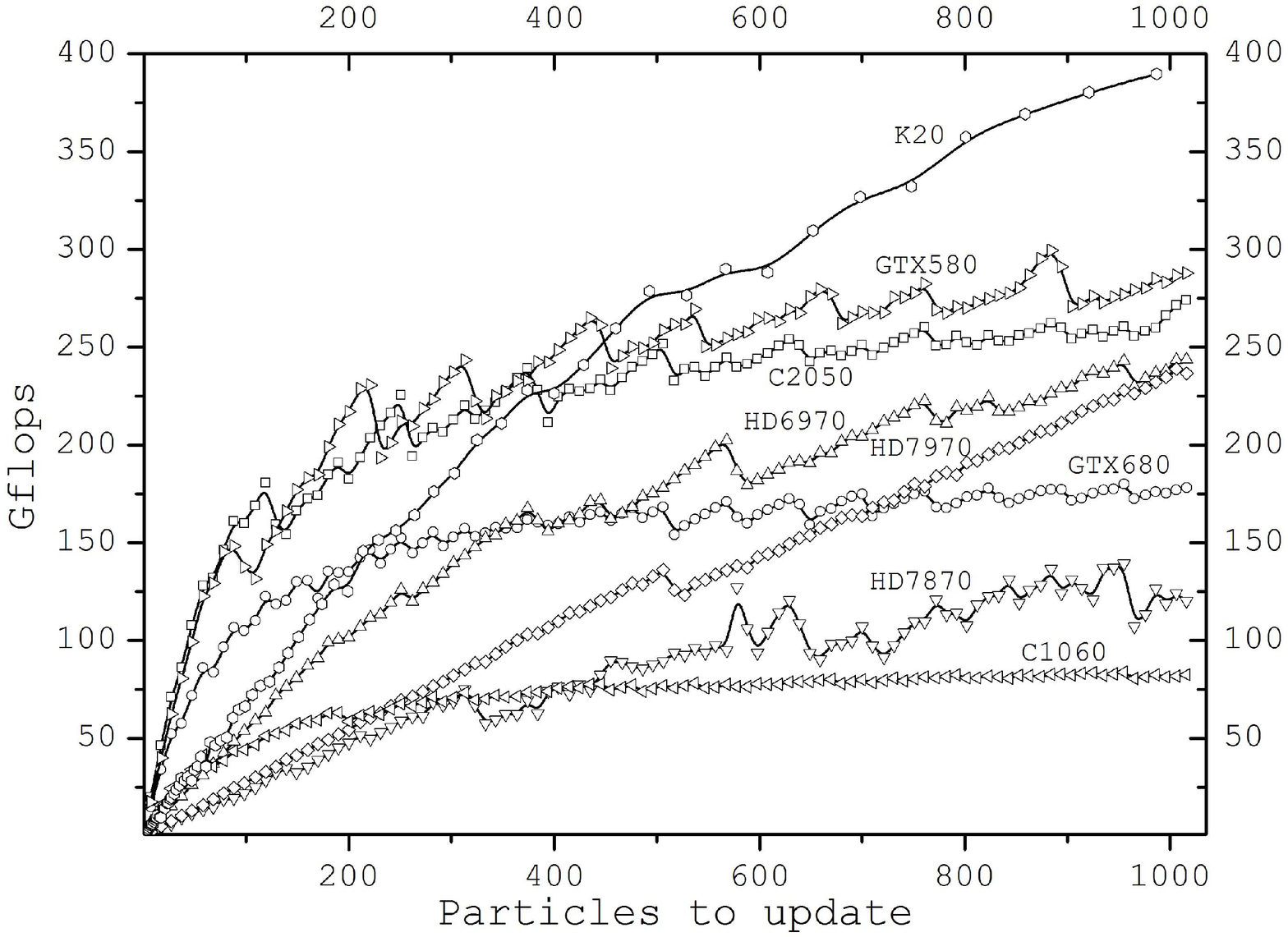}
\cprotect\caption{Speed performance of the tested GPUs, in \verb+Gflops+, as a function of the number of particles to update. This figure refers to system V3, with $N$=1,024.}
\label{fig:fig5}
\end{figure}

As we said above, and as it is shown in Fig. \ref{fig:fig3}, in this regime it is important to consider how the GPUs perform not only in the evaluation step but also in other sections of \verb+HiGPUs+. Before discussing this, let us examine the performance in the evaluation step. As we can see, for example, in Fig. \ref{fig:fig4}, relative to system \verb+V1+, the situation is completely changed. The performance of GTX 580 and Tesla C2050 becomes comparable even if they stay well below their maximum peak. On the other hand, the Radeon cards, the Tesla K20 and the old generation Tesla C1060 are slower. The growth of performance is, on average, linear for all the GPUs because they are far to be fully loaded and the performance increases with the number of running threads. In general, the larger the distance from the full-load state, the closer to the linear speed increase. This trend is particularly evident for Radeon GPUs and Tesla K20 and less for other nVIDIA cards, whose linear performance growth disappears completely already for system \verb+V3+, (see Fig. \ref{fig:fig5}). In fact, in system \verb+V3+, GTX 580, GTX 680 and Tesla C1060 get closer to their measured performance peak while Radeon GPUs and Tesla K20 maintain their approximatively linear trend being still very distant from their full load state. The Tesla C2050 performance can  further increase a little although the growth is no longer linear. We do not report further figures for the regime in which $N\in\left[1,024;16384\right]$ because the evolution of the speed performance of the tested GPUs can be argued from what has been already shown and discussed. Actually, this is a transition phase in which the situation continues to evolve and, in particular,  for $N=4,096$ Radeon GPUs and Tesla K20 have already exceeded the performance of other nVIDIA GPUs and the results become very similar to that showed in Fig. \ref{fig:mediumn}. At the light of this analysis one concludes that a GTX or a Tesla C2050 card could be the right choice to perform direct $N$-body simulations in this regime but we need to consider also other factors that will be taken into account in the next Section.

\subsection{Other important code sections}
As seen in Fig. \ref{fig:fig3}, while the evaluation section constitutes the most important part for large-$N$ systems, in the case of small-$N$ we must consider also the performance obtained in other sections, which we divide, for convenience, into 3 groups (see also Tab. \ref{tab:sections})

\begin{enumerate}
\item Host-to-Device and Device-to-Host transfers (sections 2, 7 and 10);
\item Reduction of partial forces (section 5);
\item Device-to-Device transfers (sections 6 and 11).
\end{enumerate}

It is worth noting that the last version of \verb+HiGPUs+ has the possibility to run the corrector step on the GPU. This improves performance for large$-N$ systems (especially if we run \verb+HiGPUs+ on more than one computing node) and, in addition, Sections 6 and 11 of our code are not needed anymore. Moreover, the predictor step is not considered here being always quite below the other sections. Nevertheless, to develop this work we used \verb+HiGPUs+ with the corrector performed on the CPU, and, in this case, the above listed three groups of sections contribute, with good approximation,  for about 1/3 each to the execution time not spent in the evaluation of the forces. Let us examine the performance exploited in these 3 sections.

\subsubsection{Host-to-Device and Device-to-Host Bandwidth}
\begin{figure}[!ht]
\centering
\includegraphics[width=0.8\textwidth]{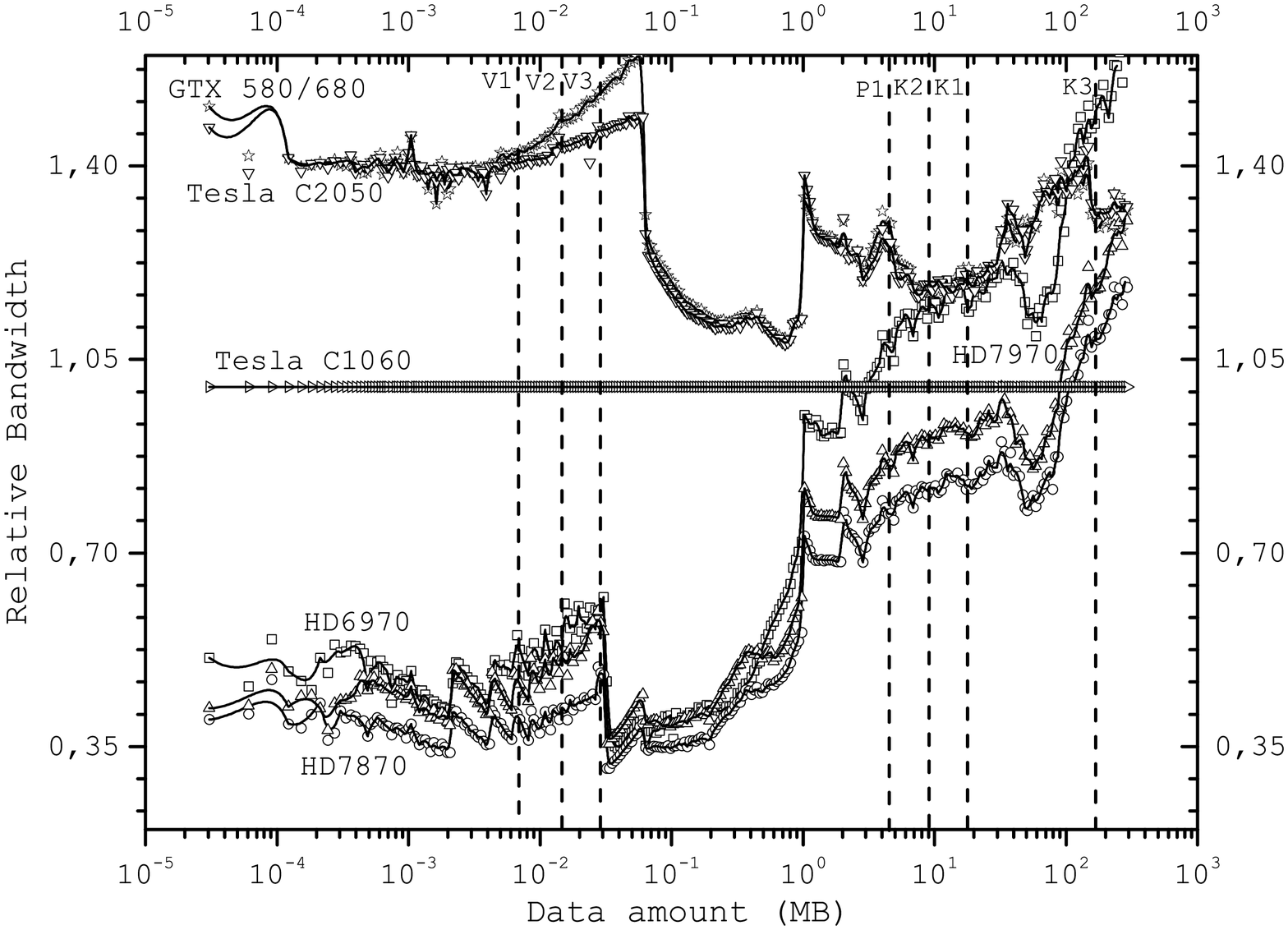}
\caption{Bandwidth, normalized to Tesla C1060, of the tested GPUs as a function of the amount of data to exchange (in MB). This figure gives also as straight vertical lines the upper limit to the amount of data that are transferred for each of the test systems. }
\label{fig:band}
\end{figure}

\begin{figure}[!ht]
\centering
\includegraphics[width=0.8\textwidth]{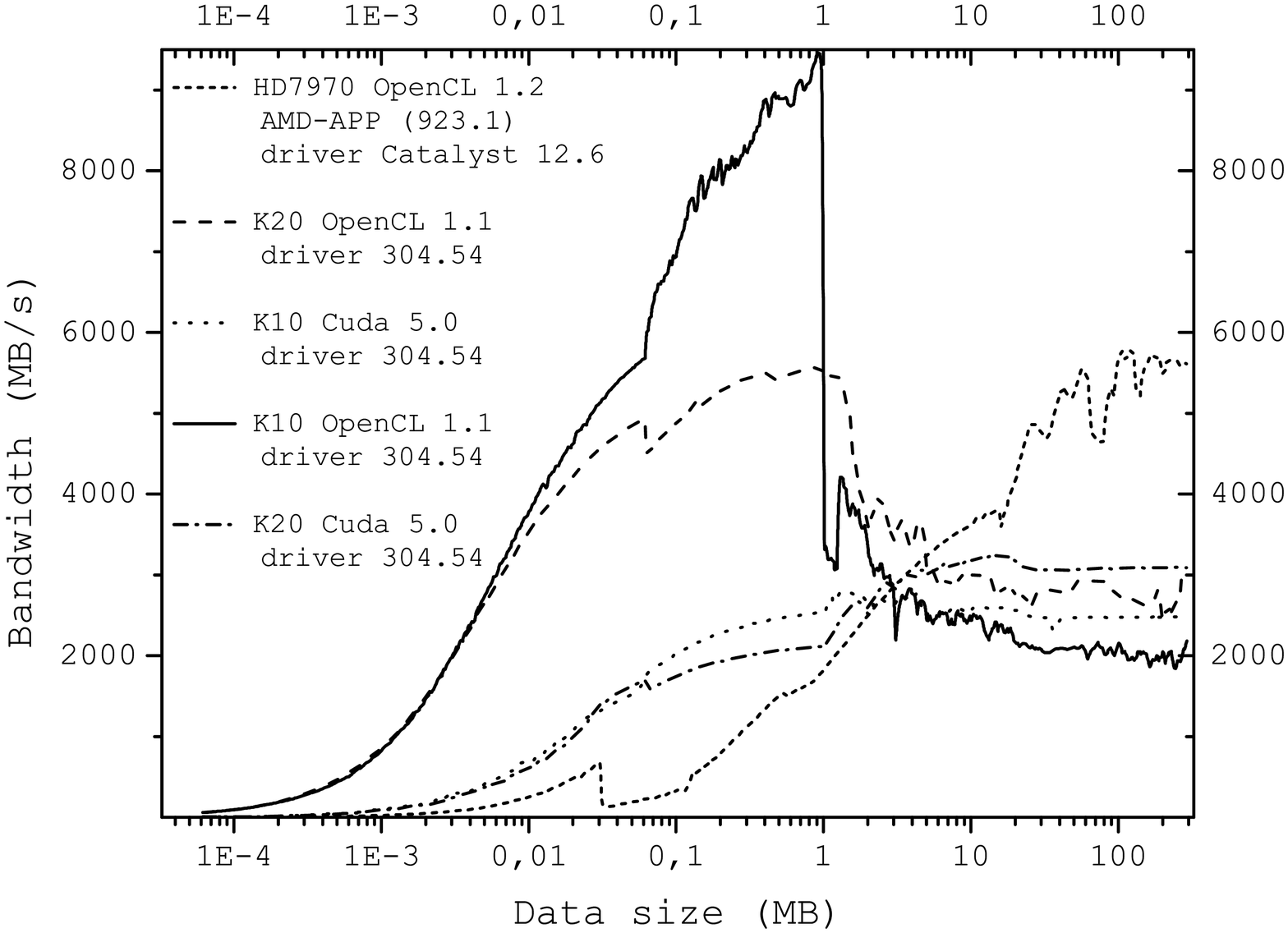}
\caption{Bandwidth of some of the different GPUs examined in this paper, in function of the amount of data transfer. In the figure we label the various GPUs with the operational software and driver version used. Note the somewhat steep decline of the bandwidth of the Tesla Kepler nVIDIA cards when using OpenCL at $\sim$1MB, while the same cards using CUDA flatten at data transfer amount above $\sim$10 MB at $\sim$3 GB/s level (that is about 50\% of the HD7970 bandwidth). }
\label{fig:k20_band}
\end{figure}

Fig. \ref{fig:band} shows the resulting bandwidth, normalized to that of Tesla C1060, in function of the amount of data transferred. The curves are obtained by an arithmetic average of the performance measured for sections 2, 7 and 10 because no significant differences were found transferring data from/to the host and device. As it can be seen in Fig. \ref{fig:band}, the results for the GTX 580 and GTX 680 are almost identical. We have also indicated, with vertical dashed lines, the maximum data transfer during the dynamical evolution of our test systems. It can be seen that the bandwidth of the Radeon GPUs is constantly below the bandwidth of nVIDIA GPUs. The reason is not easily determined but, surely, the drivers play an important role. What is important for our scopes is that this performance deficit is critical for systems \verb+V1+, \verb+V2+ and \verb+V3+ in which data transfers between the host and the device become one of the bottlenecks for our simulations. Actually, for very low-$N$ systems, Radeon GPUs loose about a factor 3.5 in performance almost independently of the number of particles in a block. This degradation of performance adds to what is lost in the evaluation step in these regimes (see Fig. \ref{fig:fig4} and Fig. \ref{fig:fig5} ). Therefore, the GTX 580/680 and the Tesla C2050 perform better also on memory transfers between host and device so they are a very good choice in regimes of weak load. Anyway the situation of weak load, i.e. low-$N$, is not in many cases critical on a computational side. Radeon GPUs improve performance when the amount of data to exchange is large enough (greater than $\sim$100 MB) but, at this level of amount of data transfer, the differences of bandwidth performance among different GPUs are definitively negligible. We do not show in Fig. \ref{fig:band} the results obtained for the Tesla K20 because we noticed that its bandwidth has a peculiar behaviour which it has been reported, for more clarity, separately in Fig. \ref{fig:k20_band}. In Fig. \ref{fig:k20_band} we report also the bandwidth of the Tesla K10, although we did not use it for performance tests, in order to point out the behaviour of a significant peak at $\sim$1MB and the following quick decrease similar to the Tesla K20. Note that the absolute values of the bandwidth depend also on the generation of PCI-Express interface, which, just in the case of the K10 GPU, is the most evolved 3.0.

\subsubsection{Reduction of partial forces}
\begin{figure}[!ht]
\centering
\includegraphics[width=0.8\textwidth]{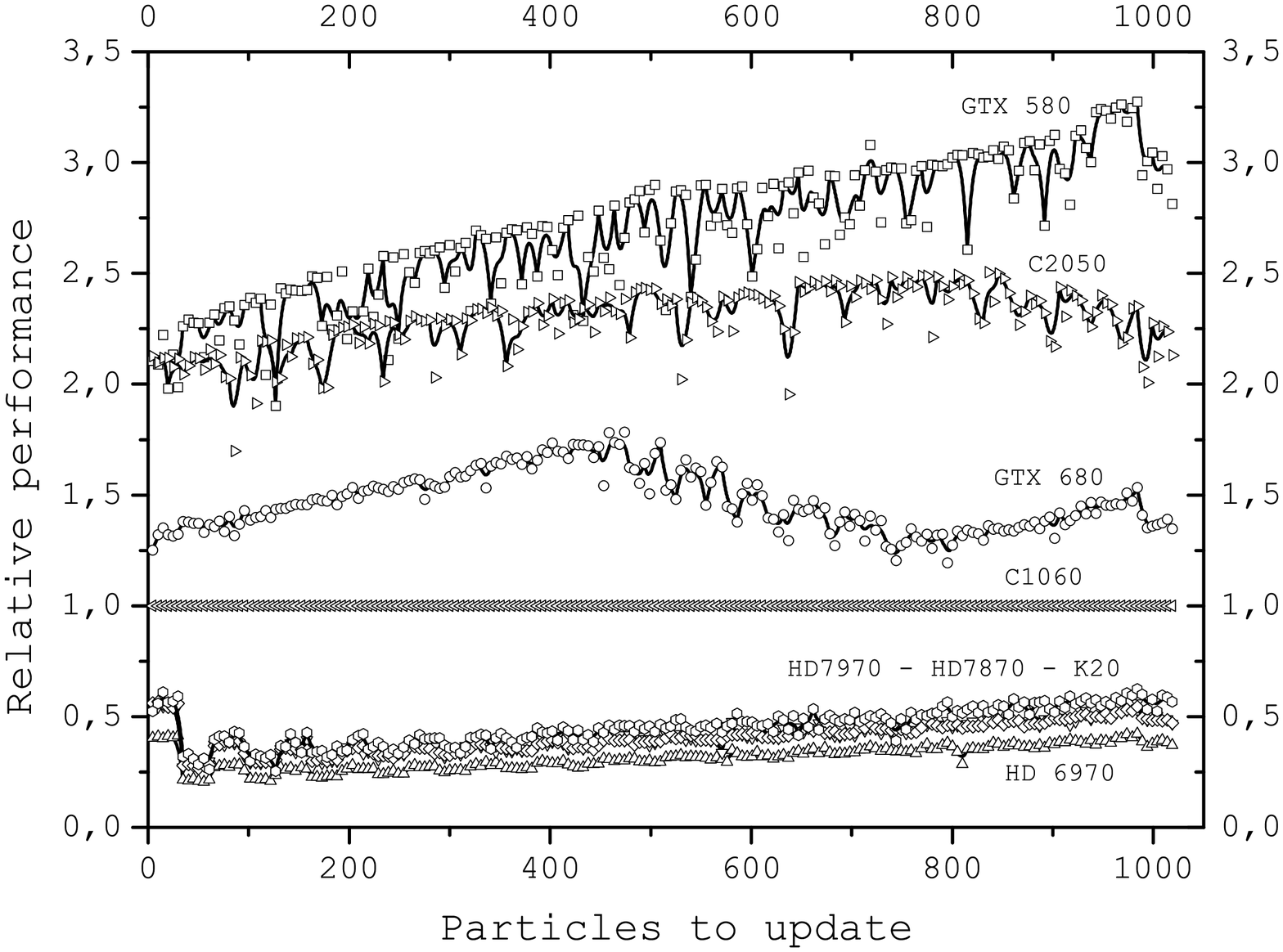}
\caption{Performance in executing the reduction of partial forces, normalized to Tesla C1060, of the tested GPUs, as a function of the particles to be updated }
\label{fig:redu}
\end{figure}

The optimizations introduced in our code are based mainly on the determination of the maximum number of threads that the GPU can handle at the same time. \verb+HiGPUs+ automatically calculates this number, $P_t$, and, if $m\lesssim P_t$, the standard one-to-one correspondence between particles to update and parallel threads is increased in order to exploit, as much as possible, all the capabilities of the GPU. In this case the correspondence is increased to 1:$k$ ($k>1$), which means that each thread calculates the force acting on its particle due to $N/k$ bodies. The performance of the evaluation step can be improved up to a factor 100 using this strategy \citep{dolcetta12}. Nevertheless, in this way we introduce another operation which is the reduction of $mk $ forces, all of them stored as double precision (64-bit) values. The latter operation becomes important for the small and very small-$N$ systems \verb+V1+, \verb+V2+ and \verb+V3+, therefore in Fig. \ref{fig:redu} we show the performance of the tested GPUs in reducing partial forces for $m<1,024$ which is the typical regime in which the above described approximation strategy is active and relevant in terms of execution time. As usual we normalize the result to one GPU (in this case we use the Tesla C1060 as reference). Similar to what previously seen, the GTX and Tesla C2050 cards perform better than Radeon and K20 cards that loose a factor $>4$ with respect, for example, to a GTX 580. We may say that, in general, GTX and Tesla C2050 GPUs are better exploited and maintain a  high efficiency on both small and large scale problems while the same cannot be said for Radeon GPUs. In fact, for these regimes of both weak load and arithmetic intensities, the single-core working frequency and lower latencies accessing GPU memory become discriminant for better and worse performance. It would be interesting to have a sort of boost of the GPU single-core frequency which should be active whenever the GPU is recognized to be not in a full-load state. This could guarantee a massively parallel GPU which would remain very efficient (like the GTX 580 for example) even for weak-load regimes.

\subsubsection{Device-to-Device bandwidth}
\begin{figure}[!ht]
\centering
\includegraphics[width=0.8\textwidth]{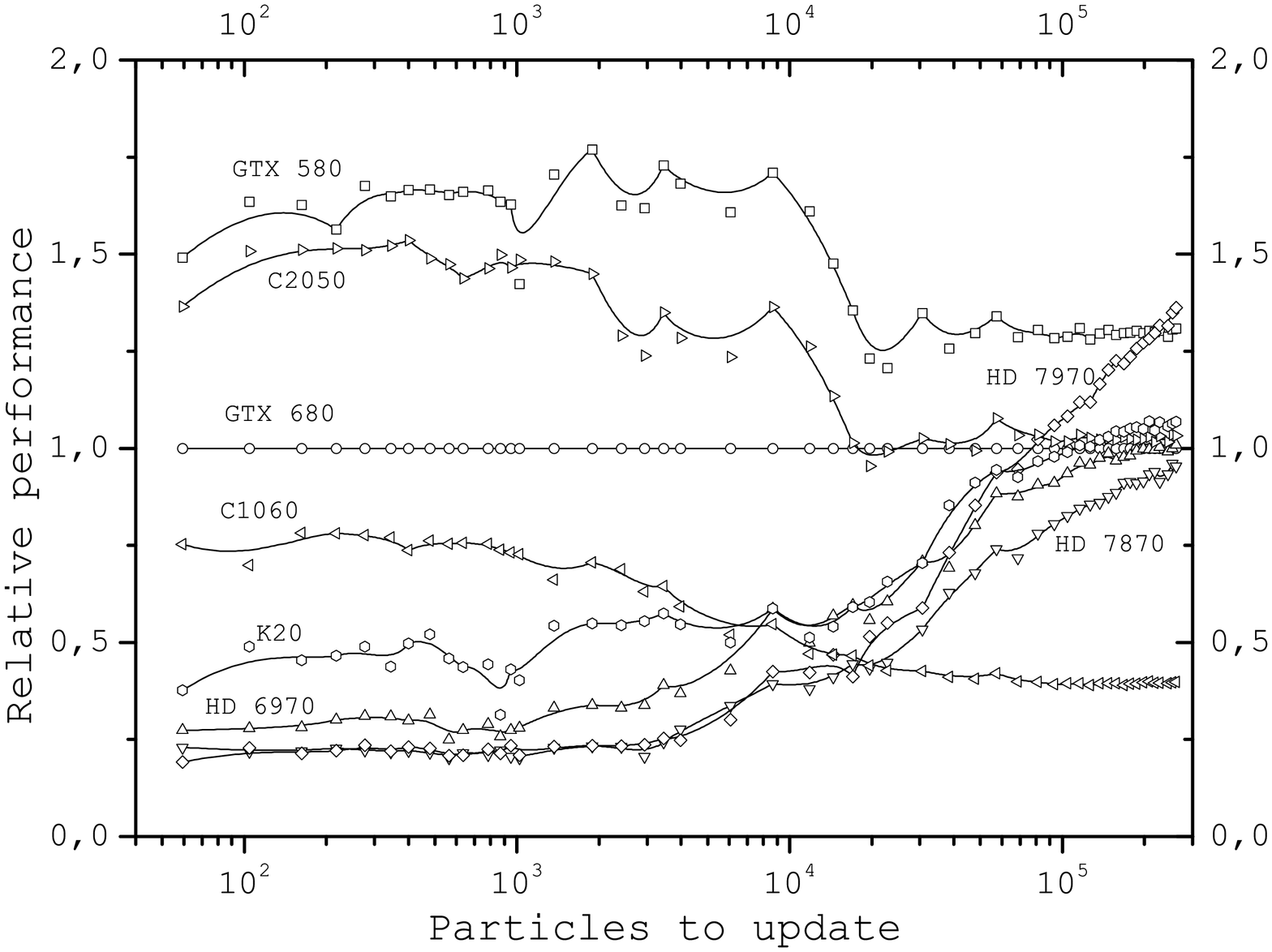}
\caption{Performance in executing device-to-device transfers, normalized to GTX 680 for convenience, of the tested GPUs, as a function of the particles to be updated }
\label{fig:d2d}
\end{figure}

For small-$N$ systems, another important section in terms of the total execution time is that involving exchanges of data inside the global memory of the single GPU. There are two kernels in \verb+HiGPUs+ which perform this sort of Device-to-Device operations, and we measured performance of these sections in terms of GB transferred per second, considering the values listed in Tab. \ref{tab:sections}. In Fig. \ref{fig:d2d} we show the results normalized, for convenience, to the performance of the GTX 680.
Once again the GTX GPUs and the Tesla C2050 are well above the Radeon GPUs, at least for $m \lesssim 10^4$. The old generation Tesla C1060 card looses a factor between 1 and 2.5 respect to the GTX 680. Radeon GPUs and Tesla K20 reach performance of the other nVIDIA cards only for $m > 10^5$; Tesla C1060 is limited by its low theoretical device-to-device bandwidth (see Tab. \ref{tab:gpu2}). Anyway, in this large $m$ regime, the difference in performance executing memory transfer operations is negligible with respect to the total execution time.
 
\section{A possible application: the Milky Way Nuclear Star Cluster}
\label{sec:exectimes}
\begin{figure}[!ht]

\includegraphics[scale=0.28]{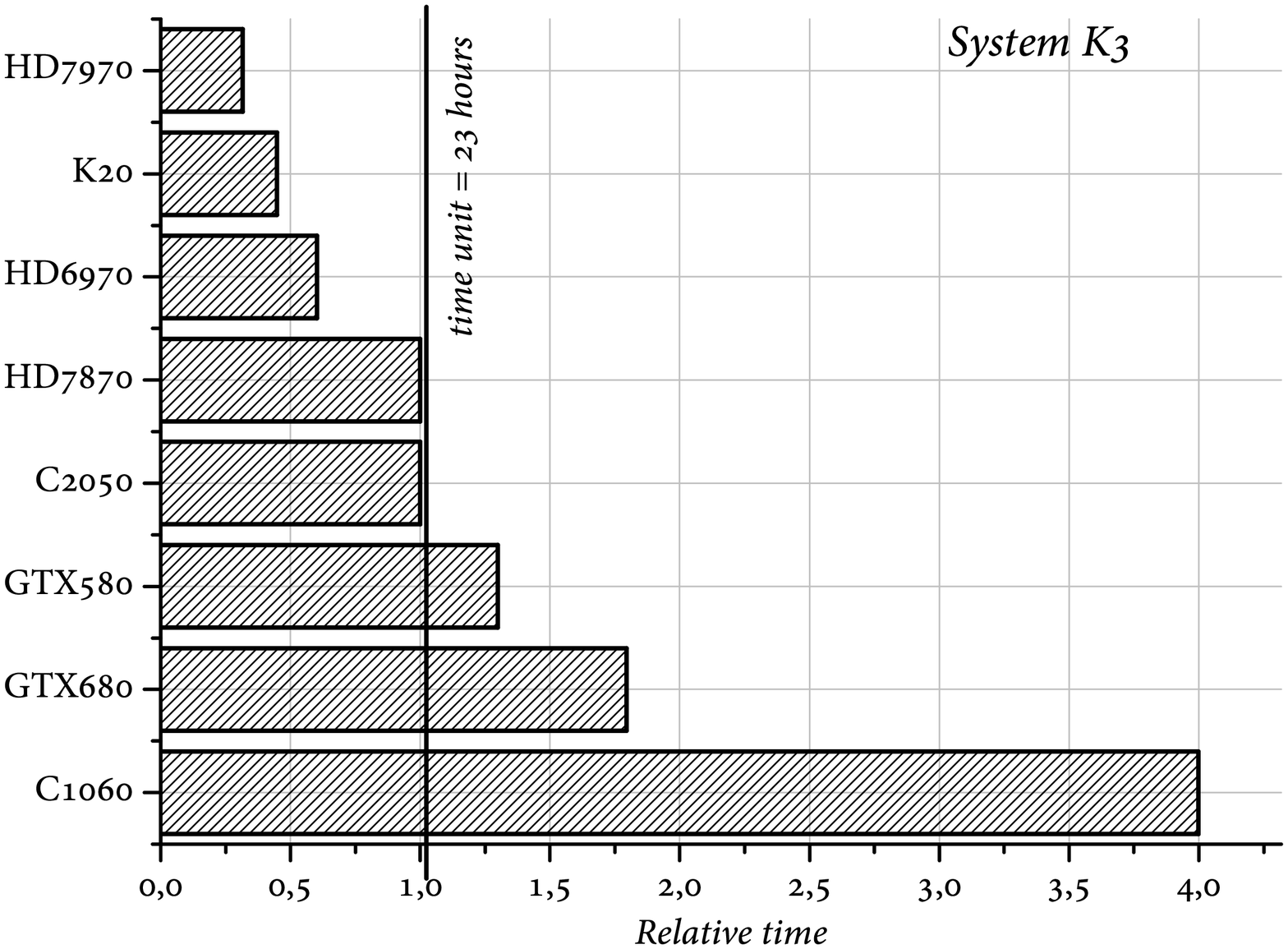}
\includegraphics[scale=0.28]{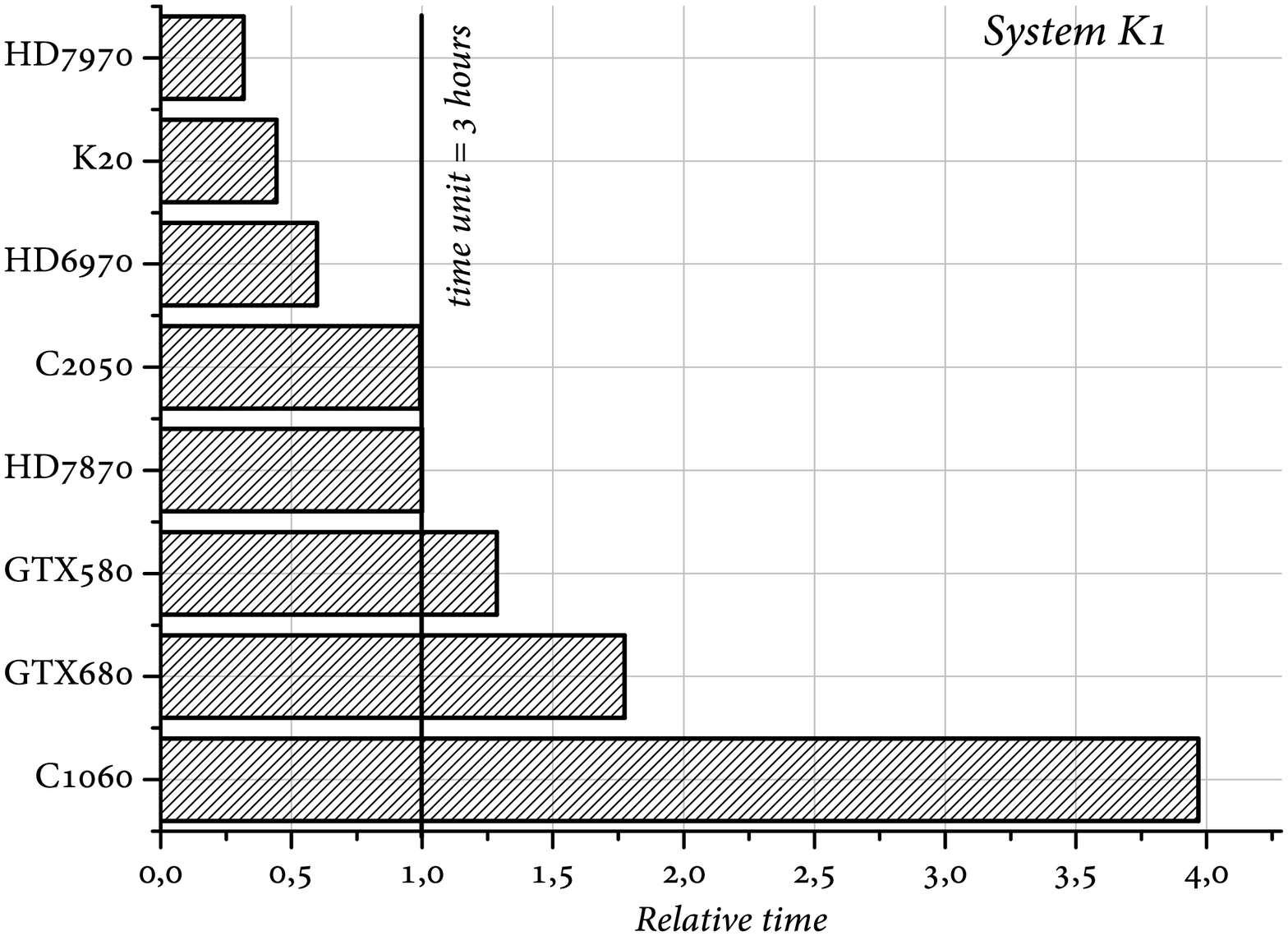}

\includegraphics[scale=0.28]{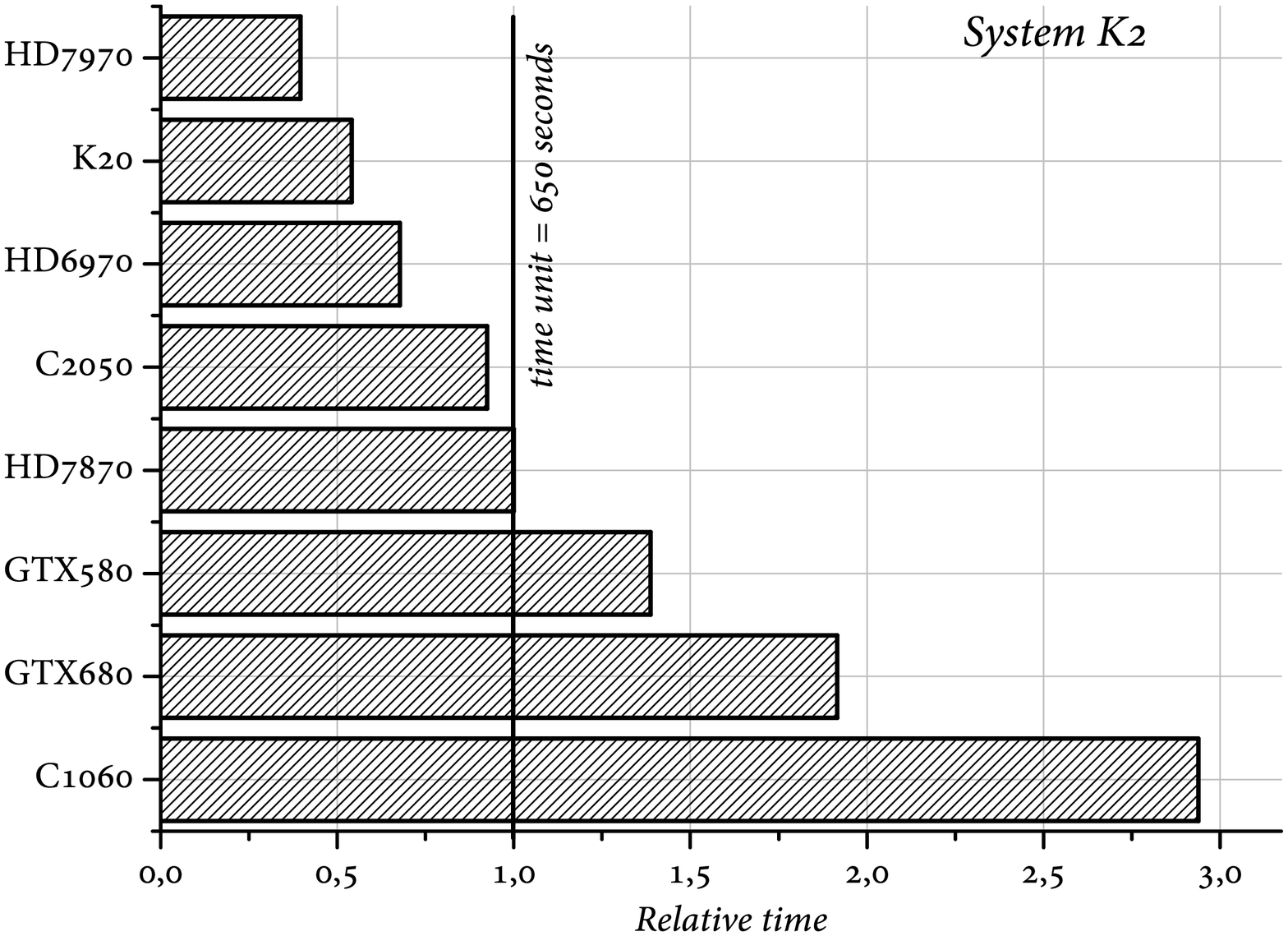}
\includegraphics[scale=0.28]{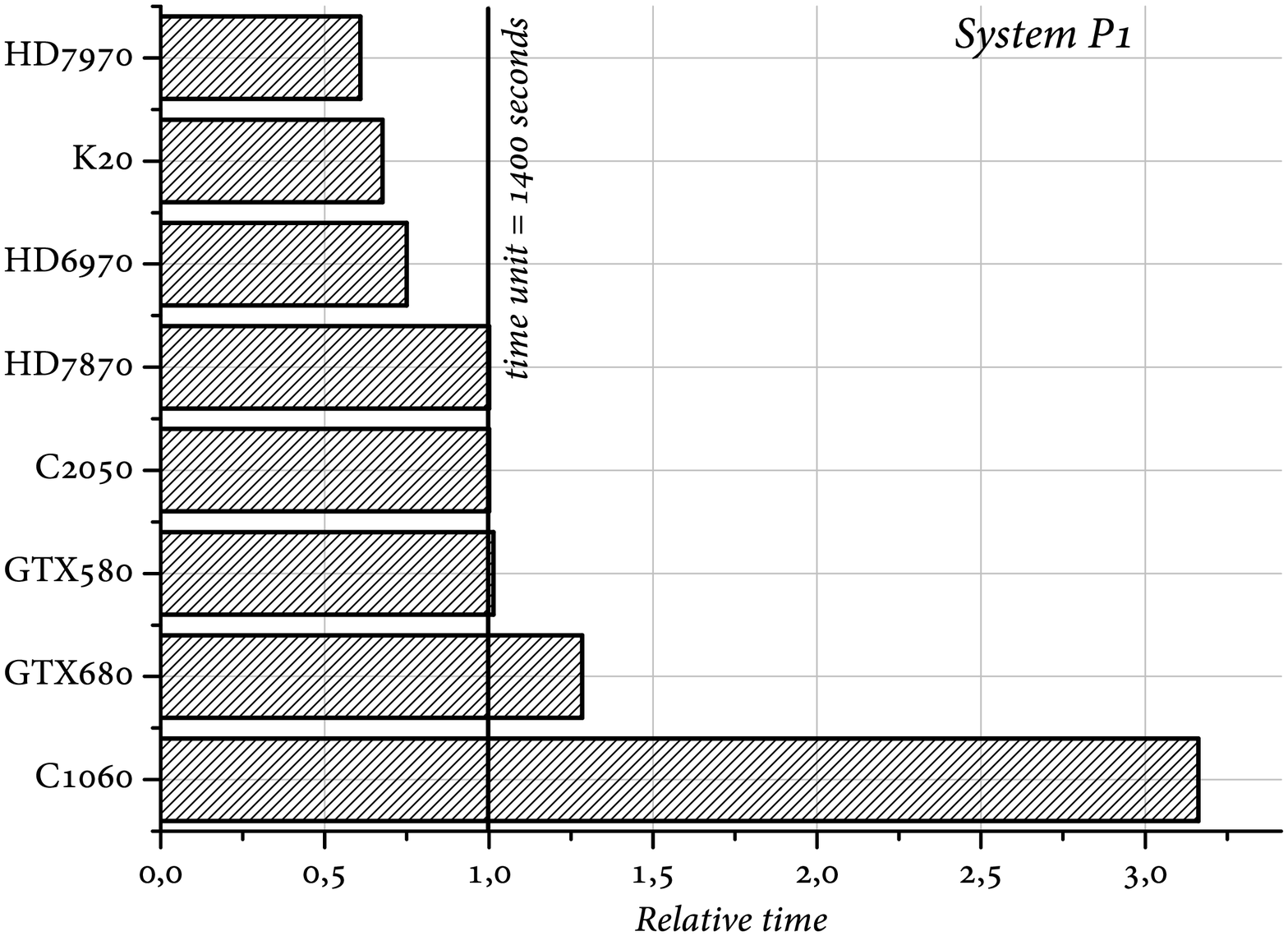}

\includegraphics[scale=0.28]{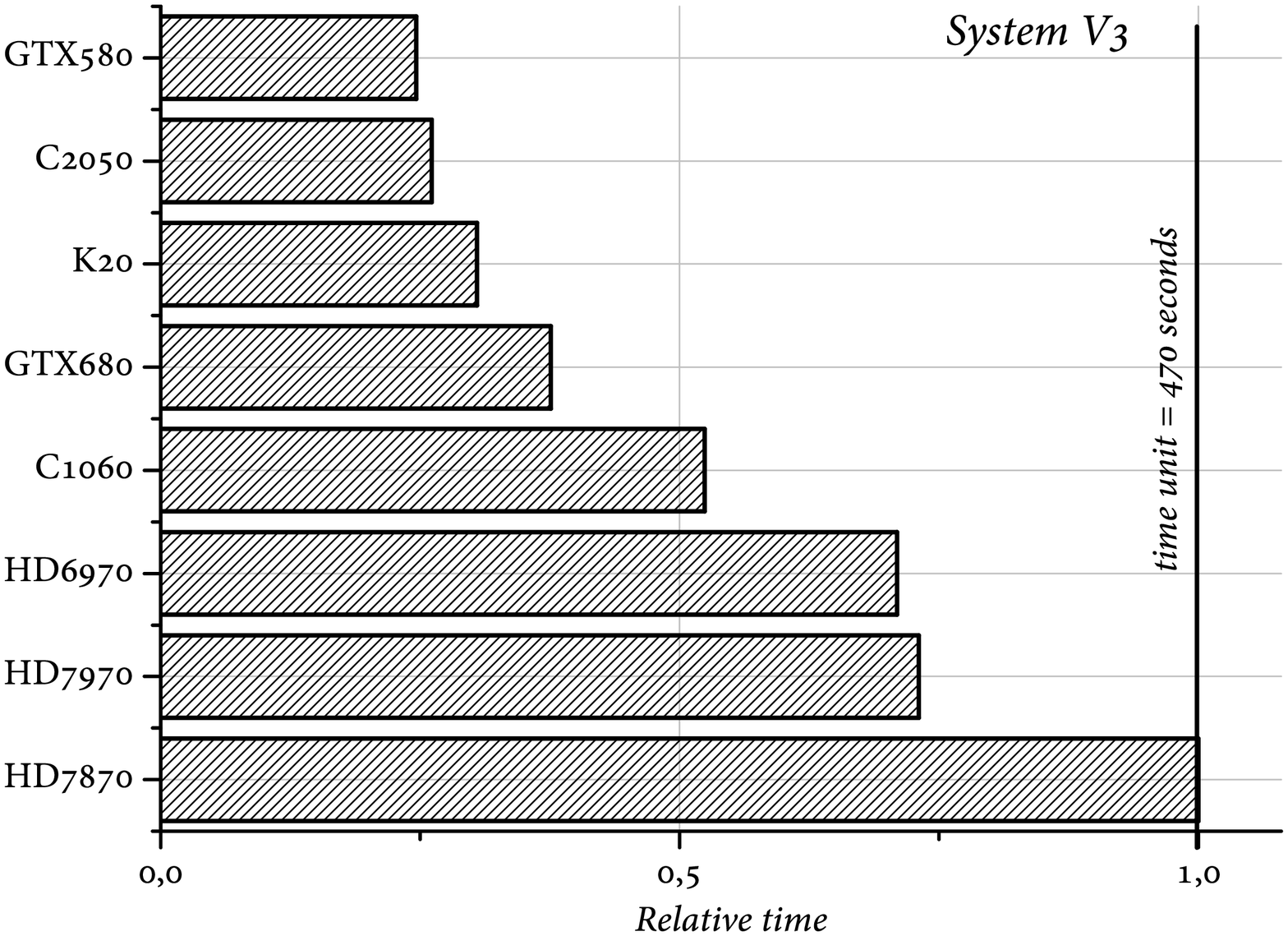}
\includegraphics[scale=0.28]{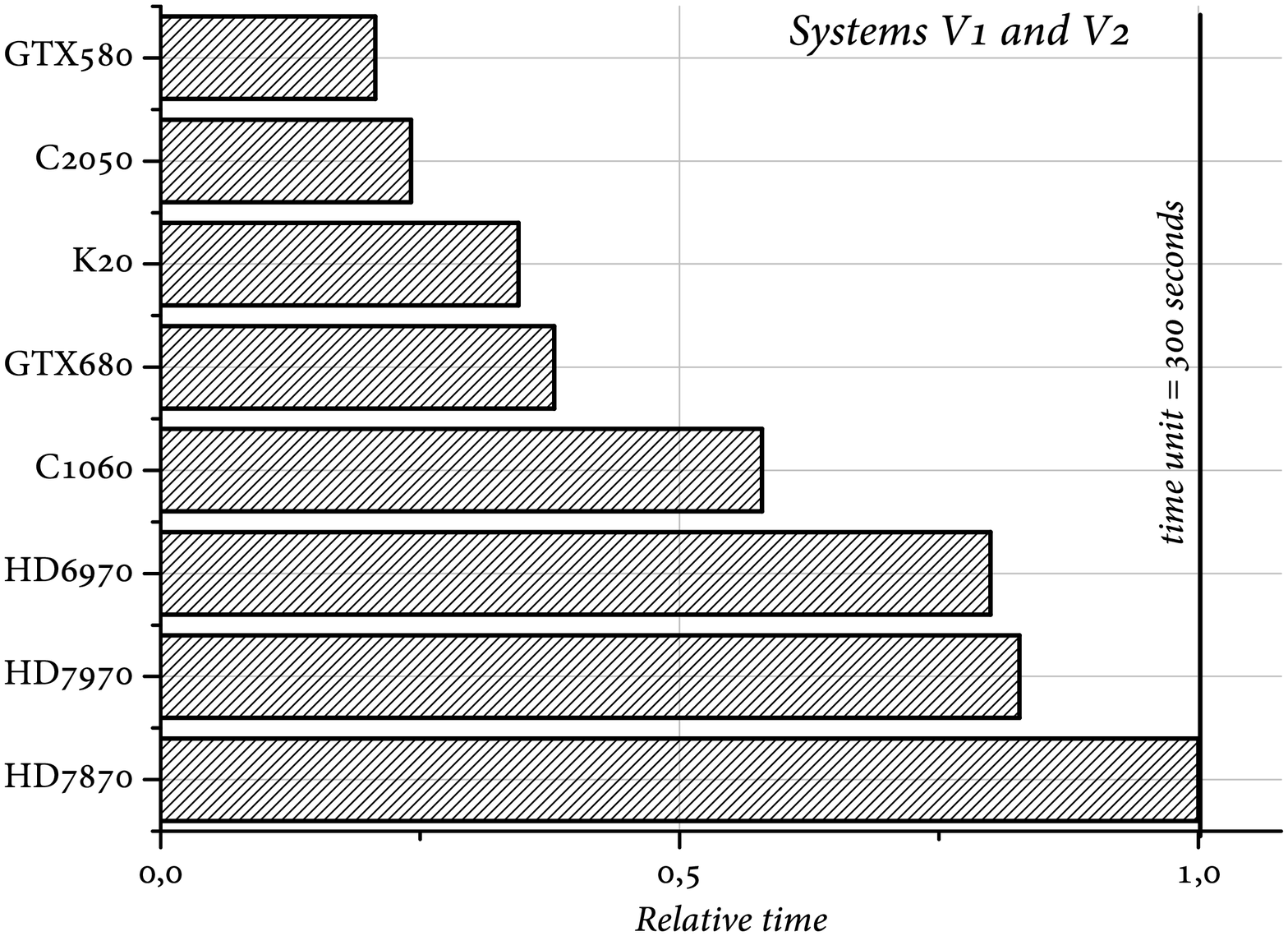}

\caption{The execution times needed to evolve the test systems over 10 time units using different GPUs. The performance are normalized to the HD 7870. The time unit is reported in each figure.}
\label{fig:isto}
\end{figure}

We briefly show in this Section the total execution times needed to evolve our test systems over 10 time units using the GPUs under test. Each system has been integrated using a proper softening parameter, $\epsilon$, in the pair-wise force. For systems \verb+V1+, \verb+V2+ and \verb+V3+, $\epsilon\simeq 3\times 10^{-4} \left\langle D \right\rangle $ where $\left\langle D \right\rangle $ is an estimate of the nearest neighbour distance i.e. 
\begin{equation}
\left\langle D \right\rangle =\frac{R}{\sqrt[3]{N}}.
\end{equation}
For systems \verb+K1+, \verb+K2+ and \verb+K3+ we used $\epsilon\simeq 10^{-2}r_c$ and, for system \verb+P1+, $\epsilon\simeq 4\times 10^{-3}b$. The results are shown in Fig. \ref{fig:isto} in the form of histograms in which the wall-clock times have been normalized to those of HD7870, for convenience. As an example, the integration of the system \verb+K3+ for $10^9$ years will require around 1,920 days using an HD7870 and only around 600 using a single HD7970. Specifically, using our very small, green and cheap cluster composed by two computational nodes each composed by two multicore CPUs and 4 HD 7970 GPUs, we may evolve the system \verb+K3+ for $10^9$ years in around 75 days of simulation, reaching a peak of 10 \verb+Tflops+ of sustained performance. For the sake of future applications of actual astrophysical interest we are dealing with the formation and the long term (Gyr) evolution of dense stellar systems around very compact and massive objects,  like black holes. Such systems are often observed in the central regions of galaxies; in particular, more steps forward are to be done in the numerical simulations of the so called Milky Way Nuclear Star Cluster, whose model of formation and evolution are still under debate (see \citep{antonini12} and \citep{bonn13}). Through preliminary tests, we estimated that we can evolve this system, modelled using $N=2M$ stars plus a central massive black hole, up to 1 Myr in around 8 hours. (That is $\sim$8,000 hours to evolve this system up to $10^9$ years). Although following a long term evolution is not possible using only eight HD7970, it can be done with our code on large hybrid supercomputers in the world (especially Titan, which is composed by 18,688 nVIDIA Tesla K20X). If we suppose, as we saw in our benchmarks, that the performance of a single K20X is slightly less than performance of one HD7970, the availability of 256 GPUs (less then 2\% of Titan), will allow to finish the mentioned simulation in one month, reaching an unprecedented spatial resolution at a sustained speed around 0.3 \verb+Pflops+, which is, definitely, a very good result for large-$N$ direct simulations.

\section{Conclusions}
\label{sec:conclu}
In this paper we compared performance of some Graphic Processing Units produced by different firms when applied to a scientific application. As test topic we chose the integration of the motion of $N$ objects interacting via pairwise Newtonian gravitational force, using a code that evaluates these forces via direct summation and performs the integration in time by mean of a Hermite's 6th order method. This code, \verb+HiGPUs+, has been developed by us and carefully tested in both its serial and fully parallel versions (CUDA or OpenCL for the GPUs and MPI and OpenMP for the part running in the host). In particular, it shows a very good scaling on hybrid platforms with a number of GPUs up to 256, the maximum allowed by the available resources. For a deep description of the code see \citep{dolcetta12}.
We made some modifications to the \verb+HiGPUs+ code to allow optimal performance comparisons in running simulations of interest in the classic $N$-body gravitational context. For instance, we allow the code to work in double precision when needed (mainly in the particle-particle distance and force evaluations) allowing single precision in all the other parts of the code where this does not cause a precision degrading. A this regard, we note that 
the relative error in total energy and angular momentum is always kept below  $5.0 \times 10^{-9}$ except for computationally critical runs (\verb+V1+, \verb+V2+ and \verb+V3+) in which it was below $1.0\times 10^{-3}$.
 
To allow a compared benchmark on GPUs of different makes a portable version of the code is needed, for the GPUs produced by nVIDIA are conveniently programmed in Compute Unified Device Architecture  (CUDA) while other vendors' GPUs do not support this paradigm and need to be programmed in OpenCL which, although ``young'' and thus not as developed as CUDA and still awkward to use, shows good efficiency. Actually, we see that \verb+HiGPUs+ running on nVIDIA GPUs of the Fermi generation has in its OpenCL version a performance about 95\% that in CUDA.
At this regard, we note that very few scientific applications have been implemented in OpenCL, so far. We are aware of another work by \citet{haus11}.
The hardware used for the comparisons presented here is a workstation with an ASUS motherboard running under Ubuntu releases of Linux, whose characteristics are reported in Table \ref{tab:astroc9} . The GPUs are of the AMD make (Radeon HD 6970, HD 7870 and HD 7970) and of the nVIDIA (GeForce  GTX 580 and GTX 680, Tesla C1060, C2050 and K20). 
The performance tests consisted of runs of \verb+HiGPUs+ with initial conditions aiming at the representation of the evolution of some stellar systems of astrophysical interest (stellar clusters) in three different ranges of the total number of interacting objects, $N$: low$-N$, intermediate and large$-N$ cases. The sense of low, intermediate and large has to be referred to the O($N^2$) complexity of the high precision computations done by direct summation. The description of the various test cases is given in Table \ref{tab:models}.

The main results obtained may be summarized as follows:

\begin{enumerate}

\item as expected, the global performance is a combination of the ``brute'' computational power of the GPU and of the host-to-device, device-to-host, and device-to-device bandwidth;

\item the highest computational speed, over 1 \verb+Tflops+, is reached by the AMD HD7970 integrating the system \verb+K3+ ($N=262,144$);

\item the AMD cards bandwidth is not as high as that of the whole set of nVIDIA GPUs tested whenever the amount of data to exchange is not over a certain threshold, over which the HD7970 performs as well as the more expensive Tesla C and Tesla K GPUs. 

\item the breakdown of nVIDIA cards performance at about 1 MB of data transfer is not completely understood. It may be due to the particular version (304.54) of the  driver used, at least for Tesla K10 and K20 cards, while it works fine for nVIDIA GTX 580 and 680 when using CUDA (OpenCL does not work properly with this driver version on GTX cards);

\end{enumerate}

The previous points imply what we have practically found and tested, i.e. that:

\begin{enumerate}

\item the global performance of the AMD HD7970 is the highest of the GPUs examined here whenever it is ``loaded'' enough to exploit its intrinsically higher computational power without penalization on the bandwidth side, thing that occurs always except for low values of $N$ ($N\lesssim 16k$);

\item the nVIDIA Tesla GPUs of the first, second (Fermi), and third (Kepler) generations perform in a range of speed from 10\% (Tesla C1060) up to 75\% (K20) that of HD7970;

\item the nVIDIA cards of the GTX family have speed performance in between the Tesla C1060 and those of Tesla C2050 and AMD HD7870, these latter being pretty similar.

\end{enumerate}

At the light of previous considerations and results, we may say that it is absolutely well pursuing code implementations in both CUDA, to exploit at best the performance of the very stable and controlled nVIDIA GPUs of the Fermi and Kepler class, and in OpenCL, which is needed to use the high power to cost ratio of the GPUs of the AMD Radeon series. With regard to these latter GPUs, we note that they are the fastest in single precision computations and, although their declared performance in double precision is lower than $1/4$ than that in single they give excellent performance when using a code which cleverly uses double precision when needed and not throughout all its run. As expected, some weak points are found in using AMD GPUs, like that of some instability as seen when using AMD different drivers on different Operating System releases. No particular problems rose, on the other side, by the absence in the Radeon and GeForce Gpus of the Error correcting code memory (ECC) available on the Tesla C2050 and K20. Also the on board memory limited to 3GB may represent, for the AMD HD7970 examined here, a limitation for some scientific applications, although it did not limited its performance in the cases studied in this paper.
The GPU hardware evolution is fast, and some developments have been announced by the GPUs producers, so no specific firm conclusion and operational suggestion may be reliably drawn to be applied over a reasonable time range. Anyway, at this stage it seems that a good receipt to follow when setting up a hybrid computational platform, especially of small-intermediate size, is to carefully consider the weights to give to the various involved parameters (cost of the single GPUs, stability and robustness of the system, quality of drivers, etc., bandwidth, power consumption on a side, performance and easiness in programming on another side) when aiming to a specific category of scientific topics.

\section{Acknowledgments}
We warmly thanks prof. S.F. Portegies Zwart and J. Bedorf at the Leiden Observatory of the Leiden University (NL) for allowing us a Tesla K20 for some of the tests reported in this paper. We also thank M. Lulli, PhD student at the Dep. of Physics of Sapienza, Univ. of Roma (Italy), for making us available the nVIDIA GTX 580 and 680 we used for the purposes of this paper.
Finally, a grateful acknowledge is also due to the E4 Computer Engineering society (Scandiano Reggio nell'Emilia, Italy) that gave us the possibility to make performance testing on one of their Tesla K10 and K20 and to S. Tinti (HPC team leader of E4) for useful discussions about bandwidth performance of both K10 and K20.

\bibliographystyle{elsarticle-num-names}
\bibliography{capuzzodolcetta}

\end{document}